\documentclass[twocolumns,structabstract]{aa}
\usepackage[latin1]{inputenc}
\usepackage[dvips]{graphicx}
\usepackage{amsmath}
\usepackage{amssymb}
\usepackage{txfonts}
\usepackage[comma,authoryear]{natbib}
\usepackage{hyperref}
\usepackage{amsmath}

\newcommand{\dpart}[2]{\frac{\partial #1}{\partial #2}}

\renewcommand{\dfrac}[2]{\frac{\mathrm{d} #1}{\mathrm{d} #2}}
\renewcommand{\l}{\ell}
\newcommand{\ie}{\textit{i.e.}}
\newcommand{\eg}{\textit{e.g.}}

\newcommand{\vect}[1]{\vec{#1}}
\newcommand{\abis}{a^{\star}}
\newcommand{\bbis}{b^{\star}}
\newcommand{\gammabis}{\gamma^{\star}}
\newcommand{\atbis}{\tilde{a}^{\star}}
\newcommand{\btbis}{\tilde{b}^{\star}}
\DeclareMathOperator*{\argmin}{arg\,min}
\DeclareMathOperator*{\argmax}{arg\,max}

\begin{document}

\titlerunning{Inequalities on stellar rotational splittings}
\title{Inequalities on stellar rotational splittings derived
       from assumptions on the rotation profile}
\author{D. R. Reese\inst{1,2}}

\institute{School of Physics and Astronomy, University of Birmingham,
           Edgbaston, Birmingham, B15 2TT, UK \\
           \email{dreese@bison.ph.bham.ac.uk}
           \and
           Stellar Astrophysics Centre (SAC),
           Department of Physics and Astronomy, Aarhus University,
           Ny Munkegade 120, DK-8000 Aarhus C, Denmark
          }

\date{}

\abstract
{A number of pulsating stars with rotational splittings have been observed
thanks to the CoRoT and \textit{Kepler} missions.  This is particularly true of
evolved (sub-giant and giant) stars, and has led various groups to
investigate their rotation profiles via different methods.}
{We would like to set up some criteria which will help us to know whether a
decreasing rotation profile, or one which satisfies Rayleigh's stability
criterion, is compatible with a set of observed rotational splittings
for a given reference model.}
{We derive inequalities on the rotational splittings using a reformulated
version of the equation which relates the splittings to the rotation profile and
kernels.}
{These inequalities are tested out on some simple examples.  The first
examples show how they are able to reveal when a rotation profile is
increasing somewhere or inconsistent with Rayleigh's criterion in a main sequence star, 
depending on the profile and the $\ell$ values of the splittings.
The next example illustrates
how a slight mismatch between an observed evolved star and a reference model can
lead to erroneous conclusions about the rotation profile.  We also show how
frequency differences between the star and the model, which should normally
reveal this mismatch, can be masked by frequency corrections for near-surface
effects.}
{}


\maketitle

\section{Introduction}

The CoRoT and \textit{Kepler} space missions have obtained exquisite pulsation
data for many stars \citep{Baglin2009, Borucki2009}. This has enabled the
detection of rotational splittings in a number of stars, including subgiants and
giants \citep{Beck2012, Deheuvels2012, Mosser2012, Deheuvels2014}, and a main
sequence star \citep{Kurtz2014}.  A number of prior studies have also used
ground-based data to extract rotational splittings. Based on these splittings,
the above authors have inverted or constrained the differential rotation
profile, and hence constrained angular momentum transport within stars.  In
particular, the core of red giants rotate much more slowly than what is expected
based on theory, thereby pointing to unknown powerful angular transport
mechanisms which operate throughout the stellar lifetime \citep{Eggenberger2012,
Marques2013, Ceillier2013}.


There are several inverse methods used to probe the internal rotation profile of
a star.  One of these, the regularised least-squares (RLS) method, searches for
the optimal profile which reproduces the observed splittings.  When carrying out
such an inversion, it is necessary to introduce \textit{a priori} information. 
Indeed, rotational splittings represent a finite number of measurements or
constraints on the rotation profile, \ie, a function defined over the interval
$[0,R]$, where $R$ is the stellar radius. As the method's name suggests, this is
typically done through a regularisation term which reduces the second order
derivative of the resultant profile. Even then, the obtained solution is not
always satisfactory.  Indeed, as can be observed in, \eg, Fig.~15 of
\citet{Deheuvels2012}, the resultant profile may change signs.  Physically, this
would correspond to a star with an ``onion'' type structure with
counter-rotating shell(s), as pointed out in \citet{Deheuvels2014}.  A second,
potentially problematic, situation is when the gradient of the rotation rate
becomes positive \citep[see, \eg, Fig.~6 of][]{Corsico2011}.  Although such a
situation can occur and has been observed both at low latitudes in the sun
\citep{Schou1998} and in a terminal age main sequence A star \citep{Kurtz2014},
it seems unlikely in many cases, especially in evolved stars which are
undergoing core contraction and envelope expansion.  Therefore, it is important
to search for inversion methods which impose a positive rotation profile and,
optionally, one which decreases outwards.  Before developing such a method,
however, it is useful to check beforehand whether such a profile is compatible
with the observations for the chosen reference model.

In the present paper, we investigate under what conditions it is possible to
obtain a decreasing rotation profile, or one which satisfies Rayleigh's
stability criterion, for a set of observed rotational splittings, and a given
reference model.  In Section~\ref{sect:theory}, we show how these assumptions on
the rotation profile lead to inequalities on the rotational splittings. 
Section~\ref{sect:application} then contains two test cases.  The
first one shows how such inequalities can detect when a rotation
profile does not obey Rayleigh's stability criterion, and the second illustrates
how these can be used to reveal a mismatch between the reference model
and the star, provided one assumes a decreasing rotation profile.  A
short discussion concludes the paper.

\section{Inequalities on rotational splittings}
\label{sect:theory}

In spherically symmetric non-rotating stars, the pulsation modes are described
by three quantum number: the radial order, $n$, which corresponds to the number
of nodes in the radial direction, the harmonic degree, $\l$, which is the total
number of nodal lines on the surface, and the azimuthal order, $m$, which gives
the number of nodes around the equator.  For each pair $(n,\l)$ there are
$2\l+1$ modes with the azimuthal order ranging from $-\l$ to $\l$. These modes
are degenerate, \ie, they all have the same frequency.  If, however, the star is
rotating, these modes will no longer be degenerate.  Furthermore, if the
rotation profile is slow and only depends on the radial coordinate, $r$, then,
based on first order effects, these modes will be evenly spaced by a quantity
known as the rotational splitting, $s_{n,\l}$.  Using the variational
principle, it is possible to derive a relation between the rotational splitting
and the rotation profile, $\Omega$ \citep[\eg][]{Aerts2010}:
\begin{equation}
s_{n,\l} \equiv \frac{\omega_{n,\l,m} - \omega_{n,\l,0}}{m} = \left( 1 - C_{n,\l} \right)
           \int_0^R K_{n,\l}(r) \Omega(r) \mathrm{d}r,
\label{eq:inverse_problem}
\end{equation}
where $\omega$ is the pulsation frequency, $C_{n,\l}$ the Ledoux constant,
$K_{n,\l}$ the rotation kernel, and $R$ the stellar radius.  In the above
formula, we have assumed that the star is viewed from an inertial frame (as
opposed to a co-rotating frame), and are making use of what could be called the
``prograde convention'', \ie, modes with positive azimuthal orders are
prograde.  If the opposite convention is used, then the splitting is defined as
$s_{n,\l} \equiv \left(\omega_{n,\l,-m}-\omega_{n,\l,0}\right)/m$.

The Ledoux constant takes on the following expression \citep{Ledoux1951}:
\begin{equation}
C_{n,\l} = \frac{\int_0^R \left(2\xi\eta + \eta^2\right)\rho r^2 \mathrm{d}r}
                {\int_0^R \left[\xi^2 + \l(\l+1)\eta^2\right]\rho r^2 \mathrm{d}r},
\end{equation}
where $\xi$ and $\eta$ are the radial and horizontal Lagrangian displacements,
respectively, and $\rho$ the density profile of the star.  Likewise, the
rotation kernel can be expressed as follows:
\begin{eqnarray}
K_{n,\l}(r) &=& \frac{\left[\xi ^2 + \l(\l+1) \eta^2 - 2\xi\eta -\eta^2\right] \rho r^2}
                     {\int_0^R \left[\xi ^2 + \l(\l+1) \eta^2 - 2\xi\eta -\eta^2\right] \rho r^2 \mathrm{d}r} \nonumber \\
            &=& \frac{\left[(\xi - \eta)^2 + (\l^2 + \l - 2) \eta^2\right] \rho r^2}
                     {\int_0^R \left[(\xi - \eta)^2 + (\l^2 + \l - 2) \eta^2\right] \rho r^2 \mathrm{d}r}.
\label{eq:rotation_kernel}
\end{eqnarray}
Bearing in mind that rotation kernels are only defined for non-radial modes $(\l
\geq 1)$, it is straightforward to see that $K_{n,\l}(r)$ is positive for all $r$
values.  Furthermore, $K_{n,\l}$ is unimodular, \ie, $\int_0^R K_{n,\l} (r)
\mathrm{d}r = 1$.

\subsection{Rotation profiles with a negative gradient}

\subsubsection{Inequalities for the full domain}
At this point, we introduce two assumptions concerning the rotation profile:
\begin{enumerate}
\item The gradient of the rotation profile is negative $(\dfrac{\Omega}{r} <
      0$).  As mentioned in the introduction, there are stars where this is not
      the case \citep{Schou1998, Kurtz2014}, but we expect this to be true in
      many cases, especially in sub-giants and giants.
\item The surface rotation rate is positive.  When combined with the previous
      assumption, this implies that the entire rotation profile is positive,
      which avoids onion-type structures with counter-rotating shells.
\end{enumerate}
We return to Eq.~(\ref{eq:inverse_problem}) and do an integration by parts of
the right hand side:
\begin{equation}
\frac{s_i}{1-C_i} = \Omega(R) - \int_0^R \left(\dfrac{\Omega}{r} \int_0^r K_i(r') \mathrm{d}r' \right) \mathrm{d}r,
\label{eq:inverse_problem_reformulated}
\end{equation}
where we have made use of the fact that $K_{n,\l}$ is unimodular, cancelled out
one of the terms, and used the index $i$ as shorthand for $(n,\l)$. If the
rotation profile has a discontinuity at $r_d$, then the integration by parts can
be carried out on the domains $[0,r_d]$ and $[r_d,R]$ separately:
\begin{eqnarray}
\frac{s_i}{1-C_i} &=& \Omega(R)
     +  \left[\Omega(r_d^-) - \Omega(r_d^+)\right] \int_0^{r_d}  K_i(r)\mathrm{d}r \nonumber \\
 & & -  \int_0^{r_d} \left(\dfrac{\Omega}{r} \int_0^r K_i(r')\mathrm{d}r' \right) \mathrm{d}r \nonumber \\
 & & -  \int_{r_d}^R \left(\dfrac{\Omega}{r} \int_0^r K_i(r')\mathrm{d}r' \right) \mathrm{d}r.
\label{eq:inverse_problem_reformulated_discontinuous}
\end{eqnarray}
We note that to be consistent with our first assumption, it makes more sense if
$\Omega(r_d^-) > \Omega(r_d^+)$.  Similar formulas can be obtained for multiple
discontinuities.  At this point, we introduce a first type of normalised
rotational splitting:
\begin{equation}
s'_i \equiv \frac{s_i}{1-C_i}.
\end{equation}

A first, and rather trivial, inequality is immediately apparent from either
Eqs.~(\ref{eq:inverse_problem_reformulated})
or~(\ref{eq:inverse_problem_reformulated_discontinuous}).  Indeed, since
$\dfrac{\Omega}{r} \leq 0$, then the last term(s) on the right-hand side is
positive.  Consequently, this leads to:
\begin{equation}
s'_i \geq \Omega(R).
\end{equation}
Such an inequality is obvious: the left-hand side is a weighted measure of the
internal rotation rate, and the right-hand side the surface rotation rate. 
Given the assumption $\dfrac{\Omega}{r} \leq 0$, the surface rotation rate is
necessarily smaller than the internal rotation rate.

\begin{figure}[htbp]
\includegraphics[width=\columnwidth]{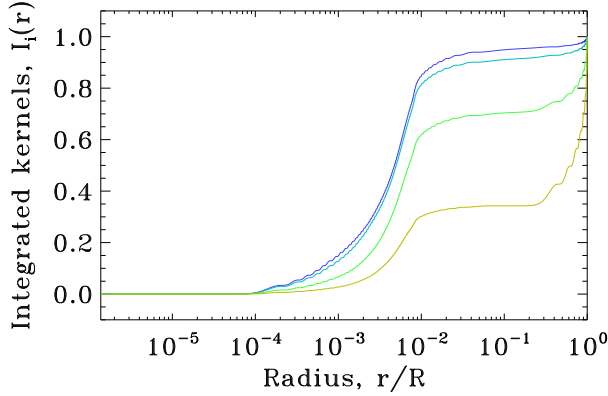}
\caption{(colour online) Integrated normalised kernels of dipole modes for a 1
$M_{\odot}$ red-giant (model 1, see Sect.~\ref{sect:application}).  The lower
curve corresponds to a more p-like mode where as the top curves are for more
g-like modes. \label{fig:integrated_kernels}}
\end{figure}

\begin{figure}[htbp]
\includegraphics[width=\columnwidth]{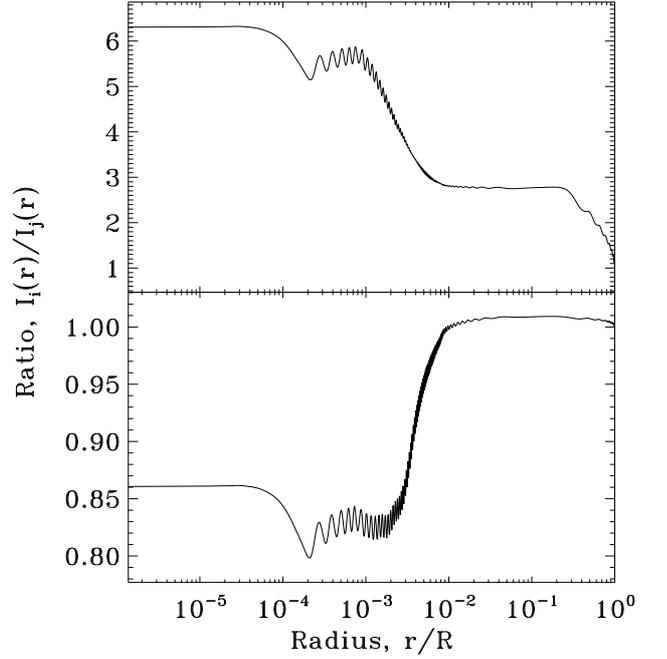} 
\caption{Ratios of integrated kernels for dipole modes.  The top panel shows the
ratio between the integrated kernels of a g-like mode and a p-like mode, whereas
the bottom panel is for two g-like modes. As can be seen in the lower panel,
the ratio crosses the value 1, which means that the two integrated kernels
have crossed.\label{fig:kernel_ratios}}
\end{figure}

\begin{figure}[htbp]
\includegraphics[width=\columnwidth]{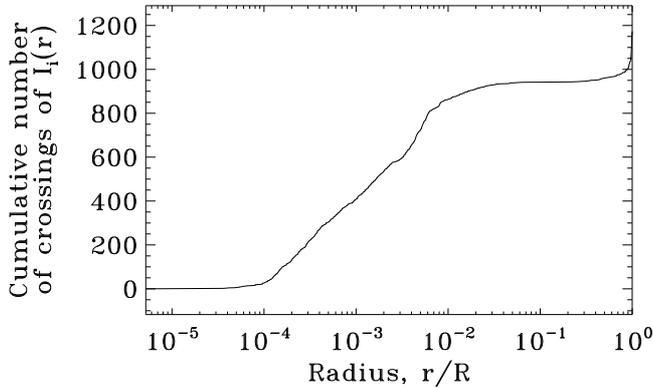}
\caption{Cumulative number of crossings between the $I_i(r)$ integrated
kernels as a function of position, $r/R$.  These crossings were calculated using
the $\l=1$ modes from Model 1, which is introduced in
Sect.~\ref{sect:red_giants}. \label{fig:crossings}}
\end{figure}

At this point, we introduce a first type of integrated rotation kernel:
\begin{equation}
I_i(r) \equiv \int_0^r K_i(r') \mathrm{d}r'.
\label{eq:integrated_kernels}
\end{equation}
Given that $K_i$ is unimodular, we have $I_i(R)=1$.  Furthermore, since $K_i$ is
positive for all $r$ and only zero in isolated points, the function $I_i$ is
strictly increasing.  Given that $I_i(0) = 0$ by construction, $I_i(r)$ is
strictly positive for $r>0$.  A visual inspection of such functions for a set
of modes, such as is illustrated in Fig.~\ref{fig:integrated_kernels}, shows that
these functions tend to ``line up'' rather than cross each other.  In other
words, if $I_i(r_0) \leq I_j(r_0)$ for a given $r_0$, where $i$ and $j$
represent two modes, then $I_i(r) \leq I_j(r)$ for all $r$ values.  A similar
behaviour certainly does not apply to the kernels themselves, hence the reason
why we work with the integrated kernels.  Now, it turns out that the integrated
kernels do cross frequently (see  Fig.~\ref{fig:crossings}), but the
general trend still leads us in the right direction.  Indeed, to make the
argument more rigorous, one simply needs to find constants, $a$ and $b$,
such that the following inequalities hold:
\begin{equation}
\forall r \in [0, R], \qquad a I_j(r) \leq I_i(r) \leq b I_j(r),
\label{eq:inequalities_integrated_kernels}
\end{equation}
The optimal values of $a$ and $b$ will then simply be:
\begin{equation}
a = \min_{r \in [0,R]} \left(\frac{I_i(r)}{I_j(r)}\right), \qquad
b = \max_{r \in [0,R]} \left(\frac{I_i(r)}{I_j(r)}\right).
\label{eq:bounds}
\end{equation}
Figure~\ref{fig:kernel_ratios} shows the ratios of integrated kernels for two
pairs of modes from which we determine $a$ and $b$.  Already, it is
straightforward to see that $a \leq 1 \leq b$ by simply inserting $r=R$ in
Eq.~(\ref{eq:inequalities_integrated_kernels}).  Furthermore, if $a = 1$ or
$b = 1$, this implies that the integrated kernels do not cross (although we do
note that by construction, they will have the same values at $r=0$ and $r=R$,
\ie\ $0$ and $1$, respectively).

The behaviour in $r=0$ is more complicated.  Indeed, one needs to consider the
limit of $I_i(r)/I_j(r)$ in Eq.~(\ref{eq:bounds}) when $r \rightarrow 0$. Hence,
it is useful to know the behaviour of the rotation kernels when $r$ goes to
$0$.  For $\l \geq 1$, the vertical and horizontal displacement behave as
$\mathcal{O}(r^{\l-1})$.  Therefore, the rotation kernels scale as $r^{2\l}$
when $r$ goes to $0$ as can be seen from Eq.~(\ref{eq:rotation_kernel}).
However, there is one exception to this rule.  Indeed, when $r$ goes to $0$, the
vertical and horizontal displacements satisfy the relation: $\xi \sim \l \eta$. 
If we replace $\xi$ by $\l \eta$ in Eq.~(\ref{eq:rotation_kernel}), then we
obtain the following expression for the numerator:
\begin{equation}
\left(2\l^2 - \l - 1\right) \eta^2 \rho r^2 = (2\l +1)(\l-1) \eta^2 \rho r^2.
\end{equation}
This expression is zero when $\l=1$.  Hence, in dipole modes, the lowest order
terms cancel out and one needs to consider higher order terms.  If we return to
expression~(\ref{eq:rotation_kernel}) and substitute $\l=1$, then the second
part of the numerator drops out and we are left with $(\xi-\eta)^2 \rho r^2$. The
next order term for $(\xi-\eta)$ behaves as $\mathcal{O}(r^2)$ since only even
powers of $r$ intervene in the displacement functions of dipole modes. When
squared and multiplied by $\rho r^2$, this leads to an $\mathcal{O}(r^6)$
behaviour, rather than the $\mathcal{O}(r^2)$ behaviour initially expected. 
Finally, if we return to the general case, the integrated kernels will simply
behave as $\mathcal{O}(r^{2\l+1})$, except for dipole modes, for which $I_i(r)
= \mathcal{O}(r^7)$.  Therefore, when considering the limit $I_i(r)/I_j(r)$, the modes $i$
and $j$ need to have the same $\l$ value or else one needs to have $\l=1$ and
and the other $\l=3$.  Otherwise, the limit will either be $0$, thereby leading
to $a=0$, or infinite, thereby leading to $b=\infty$.

If we now multiply Eq.~(\ref{eq:inequalities_integrated_kernels}) by
$-\dfrac{\Omega}{r}$ and integrate over $[0,R]$, we obtain the following
inequalities:
\begin{equation}
     a \left[ s'_j - \Omega(R) \right]
\leq   \left[ s'_i - \Omega(R) \right]
\leq b \left[ s'_j - \Omega(R) \right],
\label{eq:splittings_inequalities_original}
\end{equation}
where we have made use of the assumption $\dfrac{\Omega}{r} < 0$, and of
Eq.~(\ref{eq:inverse_problem_reformulated}), which assumes that $\Omega$ is
continuous.  Bearing in mind that $a\Omega(R) \leq \Omega(R) \leq b\Omega(R)$,
one can simplify the terms proportional to the surface rotation rate, thereby
leading us to our final set of inequalities:
\begin{equation}
a s'_j \leq s'_i \leq b s'_j.
\label{eq:splittings_inequalities_final}
\end{equation}
Although simpler, this last equation is slightly less restrictive than the
previous form.  Hence, one should use
Eq.~(\ref{eq:splittings_inequalities_original}) if the surface rotation rate is
known. If the rotation profile, $\Omega$, is discontinuous at $r_d$, it is still
possible to obtain the inequalities given in
Eq.~(\ref{eq:splittings_inequalities_final}), provided that $\Omega(r_d^-) >
\Omega(r_d^+)$, which is consistent with $\dfrac{\Omega}{r} <0$ as pointed out
earlier.  As will be discussed in Sect.~\ref{sect:errors}, observational
error bars also need to be taken into account when applying the above
inequalities.

We note, in passing, that the quantities $R-r_i$ also obey the above
inequalities, where $r_i = \int_0^R r K_i(r) \mathrm{d}r$.  Indeed, an
integration by part yields:
\begin{equation}
r_i = \int_0^R r K_i(r)\mathrm{d}r = R - \int_0^R I_i(r) \mathrm{d}r.
\end{equation}
Hence, integrating Eq.~(\ref{eq:inequalities_integrated_kernels}) over
$[0,R]$ leads to:
\begin{equation}
a (R - r_j) \leq R - r_i \leq b (R - r_j).
\end{equation}
If $a$ or $b$ is equal to $1$ for a given pair of modes (\ie\ if their
integrated kernels $I_i$ and $I_j$ do not cross, as noted above), then the slope of line
connecting $(R-r_i,s'_i)$ to $(R-r_j,s'_j)$ must be positive.  However, it turns
out that integrated kernels do cross fairly frequently (as shown in
Fig.~\ref{fig:crossings}), so one cannot rely on the slope to decide whether a
particular pair of modes obey the above inequalities.  Instead, one needs to
apply the inequalities systematically.  We nonetheless expect there to be a
general trend in an $(R-r_i,s'_i)$ diagram, based on these considerations.

Bearing in mind that the original goal, as described in the
Introduction, is to be able to find rotation profiles which are decreasing and
which do not change signs, it is interesting to look at the above problem the
other way around and see whether it is possible to construct such
rotation profiles for given rotational splittings, assuming they obey the above
inequalities.  We begin by denoting $\gamma = \frac{s'_i}{s'_j}$. 
The quantity $\gamma$ is between $a$ and $b$ since the rotational splittings
satisfy the above inequalities.  Furthermore, the function
$\frac{I_i(r)}{I_j(r)}$ ranges from $a$ to $b$, by definition of these
constants.  If this function is continuous, then there exists a point
$r_{\gamma}$ such that $\frac{I_i(r_{\gamma})}{I_j(r_{\gamma})} = \gamma$. It is
then straightforward to see that the following rotation profile reproduces the
rotational splittings and satisfies the above constraints:
\begin{equation}
\Omega(r) = \left\{
\begin{array}{ll}
\frac{s'_i}{I_i(r_{\gamma})}  & \mbox{ if } 0 \leq r \leq r_{\gamma} \\
0 & \mbox{ if } r_{\gamma} < r \leq R
\end{array}
\right..
\label{eq:solution_profile}
\end{equation}
If the function $\frac{I_i(r)}{I_j(r)}$ never takes on the value $\gamma$ due to
a discontinuity (for instance in a model with a discontinuous density profile),
then one could define a rotation profile which is discontinuous at the points
$r_1$ and $r_2$, defined such that $\frac{I_i(r_1)}{I_j(r_1)} \leq \gamma \leq
\frac{I_i(r_2)}{I_j(r_2)}$, and solve the relevant system of equations to find
by what amount the rotation profile changes at each discontinuity. A more
serious difficulty occurs if $r_{\gamma}=0$ and no alternative $r$ values could
be used to construct a solution.  In such a situation, one can only get
arbitrarily close to the solution by setting a discontinuity at $\epsilon$,
finding the corresponding rotation profile, then decreasing $\epsilon$.  Hence,
in summary, the above inequalities provide a necessary and nearly sufficient
condition on the rotational splittings of two modes for the existence of a
decreasing and positive rotation profile.

Had the above inequalities not provided a nearly sufficient condition
for the existence of such profiles, then one would be left wondering if a more
stringent set of criteria could be deduced.  However, the above results suggest
that these are the most restrictive conditions one could find for a given pair
of modes.  As we will, however, see in a later section, they do not
provide the most restrictive conditions for a set of 3 or more modes, as they
apply to 2 modes at a time.  Also, one must not forget that even if rotational
splittings do obey the above inequalities, it does not guarantee that the true
rotation profile is indeed decreasing everywhere.  Indeed, for any set of
rotational splittings, it is always possible to find rotation profiles which
have a positive gradient somewhere in the star and/or a sign change.

\subsubsection{Inequalities where the centre is excluded}
\label{sect:inequalities_no_centre}

If an upper bound on $-\dfrac{\Omega}{r}$ is known in the most central
regions of the star, then it is possible to derive slightly different
inequalities which exclude these regions and potentially lead to tighter
constraints.  We start by defining new constants, $\abis$ and $\bbis$, as
follows:
\begin{equation}
\abis = \min_{r \in [r_0,R]} \left(\frac{I_i(r)}{I_j(r)}\right), \qquad
\bbis = \max_{r \in [r_0,R]} \left(\frac{I_i(r)}{I_j(r)}\right).
\label{eq:bounds_bis}
\end{equation}
where $r_0$ is the upper bound of the central region under
consideration.  Hence,
\begin{equation}
\forall r \in [r_0,R], \qquad \abis I_j(r) \leq I_i(r) \leq \bbis I_j(r).
\label{eq:inequalities_integrated_kernels_bis}
\end{equation}
We then multiply this equation by $-\dfrac{\Omega}{r}$ and integrate
over $[r_0,R]$.  Remembering that
\begin{equation}
-\int_{r_0}^R \dfrac{\Omega}{r} I_k(r) \mathrm{d}r = s'_k - \Omega(R)
+\int_0^{r_0} \dfrac{\Omega}{r} I_k(r) \mathrm{d}r,
\label{eq:inverse_problem_reformulated_no_centre}
\end{equation}
where $k$ corresponds to $i$ or $j$, we finally obtain, after some
rearrangement and cancelling out the surface rotation terms:
\begin{eqnarray}
\abis s'_j &+& \int_0^{r_0} \dfrac{\Omega}{r}\left[\abis I_j(r) - I_i(r)\right] \mathrm{d}r \leq s'_i \nonumber \\
& \leq & \bbis s'_j + \int_0^{r_0} \dfrac{\Omega}{r}\left[\bbis I_j(r) - I_i(r)\right] \mathrm{d}r.
\label{eq:splittings_inequalities_no_centre}
\end{eqnarray}
At this point, we introduce $\mathcal{B}$, the upper bound on $-\dfrac{\Omega}{r}$
over the interval $[0,r_0]$.  The first integral in the above expression can be
bounded as follows:
\begin{eqnarray}
-\int_0^{r_0} &\dfrac{\Omega}{r}& \left[\abis I_j(r) - I_i(r)\right] \mathrm{d}r \nonumber \\
  &\leq& -\int_0^{r_0} \dfrac{\Omega}{r}\max\left[0,\abis I_j(r) - I_i(r)\right]
  \mathrm{d}r \nonumber \\
  &\leq& \mathcal{B} \int_0^{r_0} \max\left[0,\abis I_j(r) - I_i(r)\right]
  \mathrm{d}r,
\end{eqnarray}
where we have also made use of the assumption $\dfrac{\Omega}{r} \leq 0$
over the interval $[0,r_0]$. We note that, if we substitute $a$ for $\abis$,
then $\max\left[0,aI_j(r)-I_i(r)\right]$ is always zero, thereby cancelling out
the right-hand side.  The integral term would then drop out of
Eq.~(\ref{eq:splittings_inequalities_no_centre}).  A similar manipulation with
the second integral term leads to:
\begin{eqnarray}
\int_0^{r_0} &\dfrac{\Omega}{r}& \left[\bbis I_j(r) - I_i(r)\right] \mathrm{d}r \nonumber \\
  &\leq& \mathcal{B} \int_0^{r_0} \max\left[0,I_i(r) - \bbis I_j(r)\right] \mathrm{d}r.
\end{eqnarray}
Once more, if we substitute $b$ for $\bbis$, the right-had side cancels
out. When substituted into Eq.~(\ref{eq:splittings_inequalities_no_centre}), these
inequalities lead to:
\begin{eqnarray}
\abis s'_j &-& \mathcal{B} \int_0^{r_0} \max\left[0,\abis I_j(r) - I_i(r)\right] \mathrm{d}r \leq s'_i \nonumber \\
& \leq & \bbis s'_j + \mathcal{B} \int_0^{r_0} \max\left[0,I_i(r) - \bbis I_j(r)\right] \mathrm{d}r.
\label{eq:splittings_inequalities_no_centre_bound}
\end{eqnarray}
The advantage of this equation over
Eq.~(\ref{eq:splittings_inequalities_final}) is that the constants $\abis$
and $\bbis$ may be much more constraining than $a$ and $b$, given that the
latter may be overly affected by the very inner regions.  In particular, this
would allow comparisons between splittings for different $\l$ values, for which
the constants $a$ or $b$ could be $0$ or infinite, respectively.  The drawback
is trying to find an appropriate value for $\mathcal{B}$.  One can expect
$\mathcal{B}$ to become small in the central regions since $\dfrac{\Omega}{r} =
\mathcal{O}(r)$.

In much the same way as was done above, one can check to see if the
above inequalities provide a sufficient condition for obtaining a positive,
decreasing rotation profile, subject to the supplementary condition
$-\dfrac{\Omega}{r} \leq \mathcal{B}$ over the interval $0 \leq r < r_0$.  If
$\frac{\mathrm{d}\Omega}{\mathrm{d}r}$ is prescribed over the interval
$[0,r_0[$, one could apply a similar approach to what was done above and attempt
to define a radial coordinate $r_{\gammabis}$ such that such that  $\gammabis =
\frac{s'_i + \int_0^{r_0} \dfrac{\Omega}{r} I_i(r) \mathrm{d}r} {s'_j +
\int_0^{r_0} \dfrac{\Omega}{r} I_j(r) \mathrm{d}r} =
\frac{I_i(r_{\gammabis})}{I_j(r_{\gammabis})}$. If such a point exists, then it
is possible to construct a rotation profile which is decreasing and satisfies
the rotational splittings.  However,
Eq.~(\ref{eq:splittings_inequalities_no_centre_bound}) does not guarantee that
$\gammabis$ is between $\abis$ and $\bbis$, and hence that $r_{\gammabis}$
exists. Only Eq.~(\ref{eq:splittings_inequalities_no_centre}) provides such a
guarantee. One may try to adjust the function
$\frac{\mathrm{d}\Omega}{\mathrm{d}r}$ over the interval $[0,r_0[$, but would
only succeed in enforcing either $\abis \leq \gammabis$ or $\gammabis \leq
\bbis$. One could then take on a different approach, and define the points
$r_{\abis}$ and $r_{\bbis}$ such that $\frac{I_i(r_{\abis})}{I_j(r_{\abis})} =
\abis$ and $\frac{I_i(r_{\bbis})}{I_j(r_{\bbis})} = \bbis$.  As opposed to
$r_{\gammabis}$, these points are guaranteed to exist.  However, as described in
Appendix~\ref{sect:solution_no_centre},  if one defines a rotation profile with
discontinuities at those points, and which reproduces the rotational splittings,
only  Eq.~(\ref{eq:splittings_inequalities_no_centre}) guarantees that the
rotation profile will decrease across these discontinuities. Hence, only
Eq.~(\ref{eq:splittings_inequalities_no_centre}) provides a necessary and
sufficient condition for the rotation splittings to correspond to a positive
decreasing rotation profile, provided it is decreasing over the interval
$[0,r_0[$. Equation~(\ref{eq:splittings_inequalities_no_centre_bound}), in
contrast, constitutes a necessary but insufficient condition on the rotational
splittings.

\subsubsection{Three or more modes}

So far, we have only considered two modes in isolation.  However, in
typical stars, a larger number of rotational splittings are observed. Of course,
one can always apply the above inequalities to every pair of rotational
splittings. However, it is obvious that a more complicated strategy than what
was described in the previous section is needed in order to construct a rotation
profile which satisfies \textit{all} of the rotational splittings
simultaneously, and yet still has a negative gradient throughout the star. 
Indeed, certain properties of the rotation profile only emerge when a sufficient
number of rotational splittings are used together.  Hence, in what follows, we
briefly explore a way of generalising the above inequalities to more than two
modes.  Let us consider an inequality of the following form:
\begin{equation}
\forall r \in [0,R], \,\,\,\, \sum_i a_i I_i(r) \leq \sum_j b_j I_j(r),
\label{eq:inequalities_many_integrated_kernels}
\end{equation}
where we are assuming that the $a_i$ and $b_j$ are positive. Then using
the same methodology as above, we deduce the following inequalities on the
corresponding rotational splittings:
\begin{equation}
\sum_i a_i s'_i \leq \sum_j b_j s'_j.
\label{eq:inequalities_many_splittings}
\end{equation}
We note that it was possible to remove the surface rotation rate because $\sum_i
a_i \leq \sum_j b_j$ (as deduced from
Eq.~(\ref{eq:inequalities_many_integrated_kernels}) for $r=R$).  If one excludes
the central region, $[0,r_0]$, then the following inequality is obtained:
\begin{equation}
\sum_i a_i s'_i - \mathcal{B} \int_0^{r_0}
\max\left[0,\sum_i a_i I_i(r) - \sum_j b_j I_j(r)\right] \mathrm{d}r
\leq \sum_j b_j s'_j,
\label{eq:inequalities_many_splittings_no_centre}
\end{equation}
where $\mathcal{B}$ is an upper bound on $-\dfrac{\Omega}{r}$ over the
interval $[0,r_0]$.  One can also obtain the following condition by
making use of the inequalities $\int_0^{r_0} \dfrac{\Omega}{r} I_j(r)
\mathrm{d}r \leq 0$, thereby allowing us to remove these terms altogether:
\begin{eqnarray}
\sum_i a_i \left( s'_i - \mathcal{B} \int_0^{r_0}
I_i(r) \mathrm{d}r \right) \leq \sum_j b_j s'_j.
\label{eq:inequalities_many_splittings_no_centre_bis}
\end{eqnarray}
Although less restrictive than
Eq.~(\ref{eq:inequalities_many_splittings_no_centre}), this latter inequality
has the advantage of being linear with respect to the coefficients $a_i$ and
$b_j$.

The main difficulty in the above inequalities is finding the constants $a_i$ and
$b_j$ in such a way as to provide tight constraints. For instance, one could
choose a set of $b_j$ values, and search for $a_i$ values which maximise the
left hand side of Eqs.~(\ref{eq:inequalities_many_splittings})
or~(\ref{eq:inequalities_many_splittings_no_centre}) while respecting
Eq.~(\ref{eq:inequalities_many_integrated_kernels}). 
Equation~(\ref{eq:inequalities_many_integrated_kernels}) could be applied at
each mesh point, thereby providing a set of $N$ linear inequalities on the
coefficients $a_i$, where $N$ is the number of mesh points.  Maximising
the left-hand side of Eq.~(\ref{eq:inequalities_many_splittings}) would then
require methods from linear programming such as the simplex algorithm. 
A similar strategy can also be applied to
Eq.~(\ref{eq:inequalities_many_splittings_no_centre_bis}).  In contrast,
maximising the left-hand side of
Eq.~(\ref{eq:inequalities_many_splittings_no_centre}) would require a method
from mathematical optimisation, due to its non-linear form.  In the next
section, we carry out a comparison between these inequalities and inverse
methods, thereby providing further insights into how to choose the coefficients
$a_i$ and $b_j$.

\subsubsection{Link with inverse methods}
\label{sect:comparison_with_inversions}

In order to understand the link between the above inequalities and
inverse methods, we start with
Eq.~(\ref{eq:inequalities_many_integrated_kernels}) and subtract $\sum_i a_i
I_i(r)$ from both sides:
\begin{equation}
\forall r \in [0,R], \,\,\,\, 0 \leq \sum_j b_j I_j(r) - \sum_i a_i
I_i(r) \equiv \mathcal{K}(r).
\end{equation}
The right-hand side of this inequality is a linear combination of
kernels that plays exactly the same role as an averaging kernel in inverse
methods.  There is, however, one key difference\footnote{A second difference
between the averaging kernel and the function $\mathcal{K}$ is that the former
is normalised so as to yield a proper average of the rotation profile.  This,
however, is a minor issue for the inequalities presented above, since these are
not affected by the normalisation of $\mathcal{K}$.}: $\mathcal{K}(r)$ needs to
be positive everywhere, as indicated by the equation, in order for the
inequality to carry through the integration onto the rotational splittings.  In
contrast, the averaging kernels from inverse methods are not obtained with such
a goal in mind, and therefore frequently take on negative values.

This condition can be relaxed if one excludes the centre, or some other
region(s) in the star.  Indeed, in such a situation, the combined kernel,
$\mathcal{K}$, only needs to be positive in those regions which have not been
excluded.  The upper bound on $-\dfrac{\Omega}{r}$ can then be used to constrain
the supplementary terms which arise from the excluded regions, as is done, for
instance, in Eq.~(\ref{eq:inequalities_many_splittings_no_centre}).  Hence, one
could, in principle, take a set of inversion coefficients, exclude the regions
where the averaging kernel is negative, and deduce inequalities on the
rotational splittings for a given upper bound, $\mathcal{B}$.  Conversely, one
could search for the limiting value of $\mathcal{B}$ beyond which the
inequalities break down.  If the averaging kernel is well localised, one will
have primarily tested whether the rotation gradient is negative in that
particular region.  This approach could also be combined with a simplex method,
as described above, in order to find tighter constraints on the splittings.

\subsubsection{Error bars}
\label{sect:errors}

In practise, it will be necessary to take into account error bars on
observed rotational splittings.  As will be described in this section, this
introduces complications when interpreting the above inequalities.  Indeed, if
one of the inequalities is not satisfied, one can only deduce that the rotation
profile has a positive gradient with some probability that needs to be determined. 
In order to illustrate this, we start with
Eq.~(\ref{eq:inequalities_many_splittings}) as a generic form for the
inequalities, group the non-zero terms together on the left-hand side, and
introduce error terms:
\begin{equation}
\underbrace{\sum_i a_i s'_i - \sum_j b_j s'_j}_{s'} 
      + \underbrace{\sum_i a_i \varepsilon'_i - \sum_j b_j
      \varepsilon'_j}_{\varepsilon'} \leq 0,
\end{equation}
where $\varepsilon'_i$ is the error realisation on a given normalised
splitting, $s'$ represents the combined splittings, and $\varepsilon'$ the
combined errors.  For a true violation of the inequality, one needs $s' > 0$. 
However, since only the measurement $s' + \varepsilon'$ is available, one has to
evaluate the probability that $s'+\varepsilon' > \varepsilon'$.  An
obvious approach is to compare the $1\sigma$ error bar on $\varepsilon'$, which
we will denote $\sigma'$, with the measurement.  This can be obtained as a
quadratic sum of the individual $1\sigma$ error bars:
\begin{equation}
\sigma' = \sqrt{\sum_i a_i^2 (\sigma'_i)^2 + \sum_j b_j^2 (\sigma'_j)^2},
\end{equation}
where the $\sigma'_i$ are the $1\sigma$ error bars on the individual
normalised splittings.  Hence, if $s'+\varepsilon'$ is equal to $3\sigma'$, and
$\varepsilon'$ follows a normal distribution, the probability that the
inequality is violated is $99.87\,\%$ (where we've taken into account the fact
that there is violation in only one of wings of the distribution for
$\varepsilon'$).  It would be tempting to conclude that this is the probability
that the rotation profile has a positive gradient.  However, one will typically
test a large number of inequalities which increases the chances
of finding large deviations on one of the $\varepsilon'$ values and hence of
having a false alarm.  Given the correlations between the different
inequalities, it is not straightforward how to calculate the probability of a
false alarm.  As will be described in Sect.~\ref{sect:application}, we will carry
out Monte-Carlo simulations to estimate such a probability.

\subsection{Rotation profiles subject to Rayleigh's stability criterion}

\subsubsection{Inequalities for the full domain}

A different set of inequalities can be obtained from Rayleigh's stability
criterion. According to this criterion, the angular momentum must increase with
the distance to the rotation axis: otherwise, the fluid will be dynamically
unstable and can free up energy by redistributing its angular momentum
\citep[\eg][]{Rieutord1997book}.  Mathematically, this is expressed by the
condition:
\begin{equation}
\dpart{(\varpi^4\Omega^2)}{\varpi} > 0,
\end{equation}
where $\varpi$ is the distance to the rotation axis.  If we furthermore assume
that the rotation profile only depends on $r$, this
criterion then becomes:
\begin{equation}
\dfrac{(r^4\Omega^2)}{r} > 0, \qquad \mbox{\ie} \qquad \dfrac{\ln
|\Omega|}{\ln r} > -2.
\label{eq:rayleigh_criterion}
\end{equation}
Hence, this criterion gives the maximum rate at which a rotation profile can
decrease before the fluid becomes unstable.  We note that such a criterion is
incompatible with a sign change in the rotation profile, except at
discontinuities.  However, in realistic stars, viscosity, even though it is
small, would remove true discontinuities, thereby suppressing sign changes.  If
a discontinuity were present in the rotation profile, it would be stable only if
the absolute value of the rotation rate increases outward across the
discontinuity, which is the opposite from what was considered in the previous
section (if we ignore sign changes).  Finally, in real stars, this
criterion is only one of the terms in the more general Solberg-Hoiland criterion
for convective stability \citep[\eg][]{Maeder2009}.  Hence there could be
situations where an unstable rotation profile is stabilised by, say, a $\mu$
gradient.

At this point, we will assume that the rotation profile does not change
signs.  Accordingly, we will see the consequences of the inequality
$\dfrac{(r^2\Omega)}{r}>0$ rather than those of
Eq.~(\ref{eq:rayleigh_criterion}). In order to derive inequalities on the
rotational splittings, we start from Eq.~(\ref{eq:inverse_problem}), do an
integration by part, and divide both sides by $\int_0^R \frac{K_i(r)}{r^2}
\mathrm{d}r$:
\begin{equation}
\frac{s_i}{\left(1-C_i\right) \int_0^R \frac{K_i(r)}{r^2} \mathrm{d}r} = R^2\Omega(R)
                  - \int_0^R \dfrac{(r^2\Omega)}{r}
                    \left[\frac{\int_0^r \frac{K_i(r')}{(r')^2} \mathrm{d}r'}
                               {\int_0^R \frac{K_i(r')}{(r')^2} \mathrm{d}r'}\right] \mathrm{d}r.
\label{eq:inverse_problem_reformulated_r2}
\end{equation}
If a discontinuity is present, for instance at $r=r_d$, one would obtain
the following formula:
\begin{eqnarray}
\frac{s_i}{\left(1-C_i\right) \int_0^R \frac{K_i(r)}{r^2} \mathrm{d}r} &=& R^2\Omega(R)
                  + r_d^2 \left[\Omega(r_d^-) - \Omega(r_d^+)\right]
                  \frac{\int_0^{r_d} \frac{K_i(r)}{r^2} \mathrm{d}r}
                       {\int_0^R \frac{K_i(r)}{r^2} \mathrm{d}r} \nonumber \\
              & & - \int_0^{r_d} \dfrac{(r^2\Omega)}{r}
                    \left[\frac{\int_0^r \frac{K_i(r')}{(r')^2} \mathrm{d}r'}
                               {\int_0^R \frac{K_i(r')}{(r')^2}
                               \mathrm{d}r'}\right] \mathrm{d}r \nonumber \\
              & & - \int_{r_d}^R \dfrac{(r^2\Omega)}{r}
                    \left[\frac{\int_0^r \frac{K_i(r')}{(r')^2} \mathrm{d}r'}
                               {\int_0^R \frac{K_i(r')}{(r')^2} \mathrm{d}r'}\right] \mathrm{d}r.
\end{eqnarray}
Of course, the difference $\Omega(r_d^-) - \Omega(r_d^+)$ would have to
be negative, in keeping with Rayleigh's criterion.  We note, in passing, that
the rotation kernels, $K_i$, behave as $r^n$ in the centre, where $n \geq 4$. 
Hence the ratio, $K_i(r)/r^2$ always goes to zero in the centre.  From now on,
we will use the following notation to designate a second type of normalised
rotational splitting:
\begin{equation}
\tilde{s}_i \equiv \frac{s_i}{\left(1-C_i\right) \int_0^R \frac{K_i(r)}{r^2} \mathrm{d}r}.
\end{equation}

A first inequality can be found straight away, in much the same way as was done
above.  The second term on the right-hand side of
Eq.~(\ref{eq:inverse_problem_reformulated_r2}) is negative, given the assumption
$\dfrac{(r^2\Omega)}{r} > 0$.  This leads to the following inequality:
\begin{equation}
\tilde{s}_i < R^2 \Omega(R).
\label{eq:inequalities_r2_simple}
\end{equation}
This inequality sets a lower limit on the surface rotation rate based on the
internal rotation rate, as measured by the rotational splitting.  This is
consistent with Rayleigh's criterion which gives the maximum rate at which the
rotation profile can decrease.

\begin{figure}[htbp]
\includegraphics[width=\columnwidth]{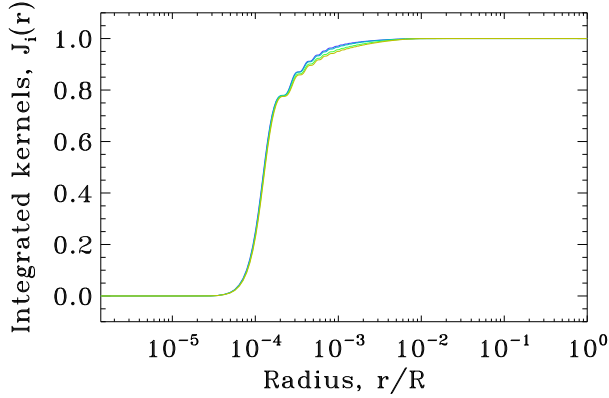}
\caption{(colour online) Second type of integrated normalised kernels for the same
modes as in Fig.~\ref{fig:integrated_kernels} (the same colour scheme is used in
both figures).  These integrated kernels are very similar, so it is difficult
to distinguish them, even though there's a mixture of p-like and g-like modes.
\label{fig:integrated_kernels_r2}}
\end{figure}

We then introduce a second type of integrated kernel:
\begin{equation}
J_i(r) \equiv \frac{\int_0^r \frac{K_i(r')}{(r')^2} \mathrm{d}r'}
                   {\int_0^R \frac{K_i(r)}{r^2} \mathrm{d}r}.
\label{eq:integrated_kernels_r2}
\end{equation}
Examples of this type of integrated kernel are shown in
Fig.~\ref{fig:integrated_kernels_r2}.

\begin{figure}[htbp]
\includegraphics[width=\columnwidth]{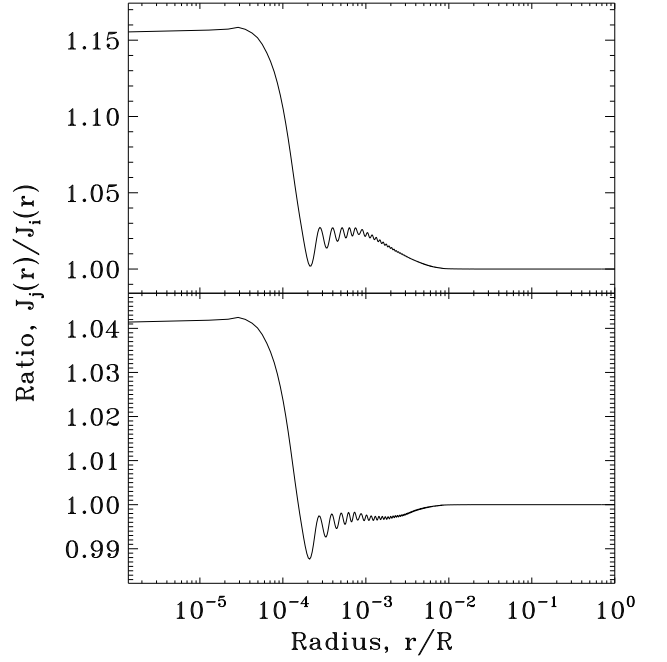}
\caption{Ratios of second type of integrated kernels for the same pairs
of modes as in Fig.~\ref{fig:kernel_ratios}.\label{fig:kernel_ratios_r2}}
\end{figure}

As was done above, we find, for a given pair of modes $(i,j)$, inequalities
of the following form:
\begin{equation}
\forall r \in [0, R], \qquad \tilde{a} J_j(r) \leq J_i(r) \leq \tilde{b} J_j(r),
\label{eq:inequalities_integrated_kernels_r2}
\end{equation}
where
\begin{equation}
\tilde{a} = \min_{r\in[0,R]} \left(\frac{J_i(r)}{J_j(r)}\right), \qquad
\tilde{b} = \max_{r\in[0,R]} \left(\frac{J_i(r)}{J_j(r)}\right).
\end{equation}
Figure~\ref{fig:kernel_ratios_r2} illustrates ratios of the second type of
integrated kernels, from which we can determine $\tilde{a}$ and $\tilde{b}$.
Once more, we obtain the relations $\tilde{a} \leq 1 \leq \tilde{b}$ by
inserting $r=1$ into Eq.~(\ref{eq:inequalities_integrated_kernels_r2}). In order
to avoid having $\tilde{a}=0$ or $\tilde{b}=\infty$, the integrated kernels
$J_i$ and $J_j$ need to have the same behaviour in the centre.  Hence, the~$\l$
values of modes $i$ and $j$, should be the same or should be~$1$ and~$3$.

We then multiply Eq.~(\ref{eq:inequalities_integrated_kernels_r2}) by
$-\dfrac{(r^2\Omega)}{r}$ and integrate over the interval $[0,R]$.   This
yields, taking into account the sign change:
\begin{equation}
     \tilde{a} \left[ \tilde{s}_j - \Omega(R) R^2 \right]
\geq           \left[ \tilde{s}_i - \Omega(R) R^2 \right]
\geq \tilde{b} \left[ \tilde{s}_j - \Omega(R) R^2 \right],
\label{eq:splittings_inequalities_r2_intermediate}
\end{equation}
where we have made use of
Eq.~(\ref{eq:inverse_problem_reformulated_r2}). This time, given the inverted
inequalities, the surface terms cannot simply be cancelled out and must
therefore be kept.  Furthermore, because of the inequality given in
Eq.~(\ref{eq:inequalities_r2_simple}), the terms in
Eq.~(\ref{eq:splittings_inequalities_r2_intermediate}) are negative.  Hence, we
rearrange this inequality so as to make positive terms appear, thereby
leading us to our final form:
\begin{eqnarray}
     \tilde{b} \tilde{s}_j - (\tilde{b}-1) \Omega(R) R^2
\leq                 \tilde{s}_i
\leq \tilde{a} \tilde{s}_j + (1-\tilde{a}) \Omega(R) R^2,
\label{eq:splittings_inequalities_r2}
\end{eqnarray}
where $\delta \tilde{s}_j$ is the error bar on the normalised splitting,
$\tilde{s}_j$. Had we worked with a discontinuous profile, the same inequalities
would be obtained, as long as the rotation rate increases across the
discontinuities (which is the opposite to what we assumed in the previous
section).  Otherwise, supplementary terms for each discontinuity where the
rotation rate decreases would need to be included, but again we emphasise that
such discontinuities would not satisfy Rayleigh's stability criterion.

\begin{figure}[htbp]
\includegraphics[width=\columnwidth]{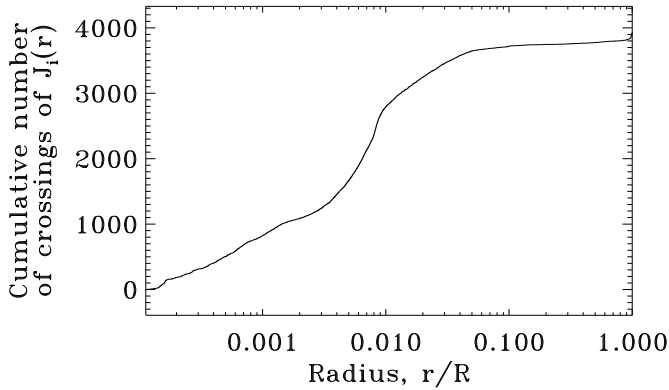}
\caption{Cumulative number of crossings of between the $J_i(r)$ integrated
kernels as a function of position, $r/R$.  These crossings were calculated using
the $\l=1$ modes from Model 1, which is introduced in
Sect.~\ref{sect:red_giants}. \label{fig:crossingsR2}}
\end{figure}

In analogy with what was done above, we introduce the quantity
$\tilde{r}_i$, defined as follows:
\begin{equation}
\tilde{r}_i = \frac{\int_0^R r \frac{K_i(r)}{r^2}\mathrm{d}r}
                   {\int_0^R \frac{K_i(r)}{r^2} \mathrm{d}r}.
\end{equation}
Carrying out an integration by parts leads to the following equality:
\begin{equation}
\tilde{r}_i = R - \int_0^R J_i(r) \mathrm{d}r.
\label{eq:r2_i}
\end{equation}
Combining Eq.~(\ref{eq:r2_i}) with the inequalities in
Eq.~(\ref{eq:inequalities_integrated_kernels_r2}) then yields:
\begin{equation}
\tilde{a} (R - \tilde{r}_j) \leq R - \tilde{r}_i \leq \tilde{b} (R - \tilde{r}_j).
\end{equation}
Hence, the quantities $R-\tilde{r}_i$ obey the \textit{opposite}
inequalities to the normalised splittings $\tilde{s}_i$.  Accordingly, if
$\tilde{a}=1$ or $\tilde{b}=1$ (\ie\ if the integrated kernels $J_i(r)$ and
$J_j(r)$ do not cross), then the slope of the line connecting
$(R-\tilde{r}_i,\tilde{s}_i)$ to $(R-\tilde{r}_j,\tilde{s}_j)$ is negative. 
However,  the integrated kernels do cross frequently, as illustrated in
Fig.~\ref{fig:crossingsR2}, so one should instead look at the general trend in
an $(R-\tilde{r}_i,\tilde{s}_i)$ diagram.  Only a systematic application of the
inequalities will yield conclusive results.

One can then try to see if the above conditions are sufficient for the
existence of a rotation profile which satisfies Rayleigh's criterion. This time,
in keeping with the inequalities, we must assume the surface rotation rate is
imposed.  As was done previously, we introduce the ratio $\tilde{\gamma} =
\frac{R^2 \Omega(R) - \tilde{s}_i}{R^2 \Omega(R) - \tilde{s}_j}$ and its
corresponding radial coordinate, $r_{\tilde{\gamma}}$, defined such that
$\tilde{\gamma} = \frac{J_i(r_{\tilde{\gamma}})}{J_j(r_{\tilde{\gamma}})}$. This
radial coordinate is guaranteed to exist provided the function
$\frac{J_i(r)}{J_j(r)}$ is continuous.  We then define a rotation profile which
is discontinuous at $r_{\tilde{\gamma}}$ and where $\dfrac{(r^2\Omega)}{r} = 0$
everywhere else:
\begin{equation}
\Omega(r) = 
\left\{
\begin{array}{ll}
\frac{\mu}{r^2}  & \mbox{ if } 0 \leq r \leq r_{\tilde{\gamma}} \\
\frac{\mu+\lambda}{r^2} & \mbox{ if } r_{\tilde{\gamma}} < r \leq R
\end{array}
\right..
\label{eq:solution_profile_r2}
\end{equation}
The constants $\mu$ and $\lambda$ are obtained by imposing the
correct surface rotation rate and rotational splittings.  This leads to:
\begin{equation}
\lambda = \frac{R^2\Omega(R) - \tilde{s}_i}{J_i(r_{\tilde{\gamma}})}, \qquad
\mu = R^2\Omega(R) - \frac{R^2\Omega(R) - \tilde{s}_i}{J_i(r_{\tilde{\gamma}})}.
\end{equation}
It is straightforward to show that $\lambda$ is positive, using
Eq.~(\ref{eq:inequalities_r2_simple}).  However, it is not clear whether $\mu$
is positive or not, meaning the above rotation profile might be negative in the
centre.  Also, we were not able to show that $|\mu| < |\mu+\lambda|$ which is
needed to ensure that the absolute value of $\Omega$ increases across the
discontinuity, and hence that Rayleigh's criterion is verified.  Instead, the
above rotation profile only satisfies the criterion $\dfrac{(r^2\Omega)}{r} \geq
0$, which is slightly less restrictive than the condition we obtained when we
assumed the rotation profile doesn't change signs in addition to Rayleigh's
criterion.  Accordingly, if $\mu$ is positive, then $|\mu| \leq |\mu+\lambda|$
and Rayleigh's criterion is \textit{nearly} satisfied (\ie\
$\dfrac{(r^4\Omega^2)}{r} \geq 0$). Hence, this simple strategy is unable to
produce a fool-proof solution solely based on the above inequalities and seems
to indicate that these do not provide a sufficient, or nearly sufficient,
condition for the existence of a rotation profile which satisfies Rayleigh's
criterion.

\subsubsection{Inequalities where the centre is excluded}
\label{sect:inequalities_r2_no_centre}

If an upper bound on $\dfrac{(r^2\Omega)}{r}$ in the central regions is
known, then it is possible to derive some slightly different inequalities, much
like in the previous section.  We begin by introducing the quantities $\atbis$
and $\btbis$:
\begin{equation}
\atbis = \min_{r \in [r_0,R]} \left(\frac{J_i(r)}{J_j(r)}\right), \qquad
\btbis = \max_{r \in [r_0,R]} \left(\frac{J_i(r)}{J_j(r)}\right).
\end{equation}
Then, following a similar reasoning as in the previous section, we
finally obtain:
\begin{eqnarray}
&\btbis& \tilde{s}_j - (\btbis\!-\!1) \Omega(R) R^2 
     - \tilde{\mathcal{B}} \!\! \int_0^{r_0} \!\!\!\!\! \max \left[0, J_i(r) - \btbis J_j(r)\right] \mathrm{d} r \nonumber \\
&\leq&                 \tilde{s}_i \nonumber \\
&\leq& \atbis \tilde{s}_j + (1-\atbis) \Omega(R) R^2 
     + \tilde{\mathcal{B}} \!\! \int_0^{r_0} \!\!\!\!\! \max \left[0, \atbis J_j(r) - J_i(r)\right] \mathrm{d} r,
\label{eq:splittings_inequalities_r2_no_centre}
\end{eqnarray}
where $\tilde{\mathcal{B}}$ denotes the upper bound on
$\dfrac{(r^2\Omega)}{r}$ over the interval $[0,r_0]$.  We do note that to first
order, $\dfrac{(r^2\Omega)}{r}$ behaves as $2r\Omega$ in the centre.

\section{Applications}
\label{sect:application}

\subsection{Direct application of the inequalities}
\label{sect:forward}

A natural and straightforward application of the above inequalities is to see
whether the rotation profile of a given star has a positive gradient and/or
fails to respect Rayleigh's criterion.  In what follows, we will therefore test
these inequalities in artificial situations to see how effective they
are.

\begin{figure}[htbp]
\includegraphics[width=\columnwidth]{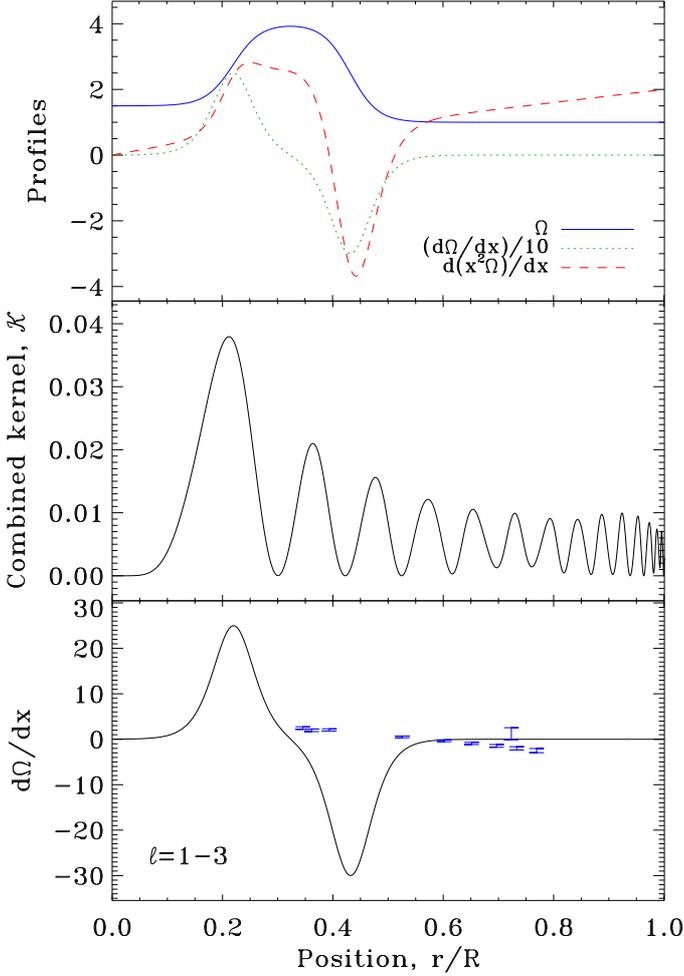}
\caption{(Colour online)  (Upper panel) Rotation profile and associated
derivatives used in first test case on Model S.  (Middle panel) Combined kernel
from simplex method which led to a detection of a positive gradient.  (Bottom
panel) SOLA inversion results on $\dfrac{\Omega}{x}$ along with $1\sigma$
error bars. \label{fig:simplex_model_S}}
\end{figure}

We start by using Model S \citep{Christensen-Dalsgaard1996} and impose the
following rotation profile:
\begin{equation}
\Omega(x) = A + B\tanh\left(\frac{x-x_0}{d}\right) 
              - C\tanh\left(\frac{x-x_1}{d}\right),
\end{equation}
where $A = B = 1.25$, $C = 1.5$, $x_0 = 0.220$, $x_1 = 0.432$, and $d = 0.05$. 
This rotation profile and its derivatives are displayed in the upper
panel of Fig.~\ref{fig:simplex_model_S}.  As can be seen, it has a
positive gradient in the first $33\%$ of the star, and does not satisfy
Rayleigh's criterion over a narrow region around $0.45\,R$.  We then considered
a set of $44$ modes with harmonic degrees $\l=1-3$ from which we calculated
``observed'' rotational splittings.  We assumed these splittings follow a normal
distribution, and their $1\sigma$ error bars are $0.0067\,\mu$Hz, which is
somewhat smaller than what is achieved with the Kepler satellite.

A systematic application of the
inequalities~(\ref{eq:splittings_inequalities_final})
and~(\ref{eq:splittings_inequalities_no_centre_bound}) to all pairs of modes
with degrees $\l=1$ to $3$ did not reveal a negative gradient.  However, using
Eq.~(\ref{eq:inequalities_many_splittings}), we were able to place tighter
constraints on the splittings.  We systematically applied this equation, in
which each splitting was isolated to one side, then the other, of the
inequality, and a simplex algorithm\footnote{The simplex algorithm was
downloaded from
\url{http://algs4.cs.princeton.edu/65reductions/Simplex.java.html} and is
described in \citet{Sedgewick2011}.} used to optimise the coefficients of the
remaining splittings.  This approach allowed us to deduce lower and upper bounds
on each splitting, and to show that the rotational splitting of the $\l=3$ mode
with the lowest radial order ($n=13$) is approximately $4.99\sigma$ beyond the
bound predicted by  Eq.~(\ref{eq:inequalities_many_splittings}), \ie\ we showed
that $s'/\sigma'=4.99$.  A naive interpretation of this would lead to the
conclusion that the probability of a false alarm is $3.0 \times 10^{-7}$. 
However, as discussed in Sect.~\ref{sect:errors}, this does not take into
account the fact that we are testing multiple inequalities, which increases the
probability of having a large deviation on one of them.  We therefore carried
out a Monte Carlo test with $10^6$ realisations of the splittings.  We
calculated the maximal value of $\varepsilon'/\sigma'$ for each realisation,
using the coefficients from the inequalities based on the mode pairs and based
on the simplex approach, and counted the number of times the threshold $4.99$
was exceeded.  This occurred in 166 cases, thereby implying a probability of
$1.7 \times 10^{-4}$ for a false alarm, \ie\ substantially larger than the above
value.  A similar test, using splittings from a decreasing profile, yielded a
probability of $1.5\times 10^{-4}$, thereby showing the result is robust. Even
then, this new value is likely to be an underestimate, because the Monte Carlo
simulation was carried out for fixed coefficients.  Ideally, one would want to
repeat the simplex calculation for each realisation of the splittings, since the
results are optimised for the given set of splittings. However, the numerical
cost of such an approach is too high to obtain a result in a reasonable amount
of time.

\begin{figure}[htbp]
\includegraphics[width=\columnwidth]{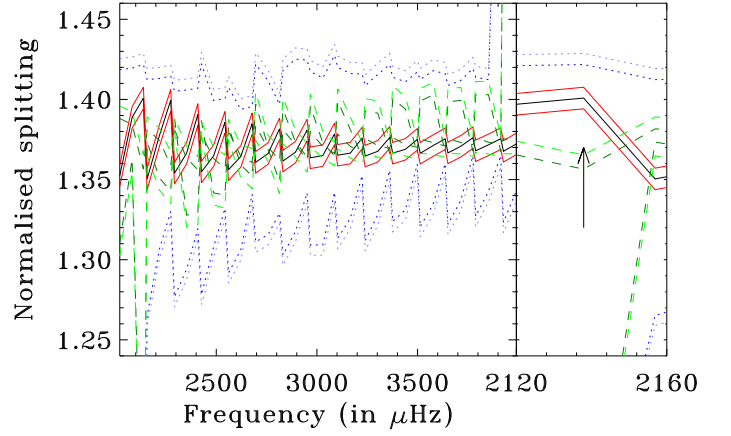}
\caption{(Colour online) Rotational splittings and $1\sigma$ error bars
(solid red lines), constraints deduced from inequalities on pairs of modes
(dotted blue lines), and constraints from multiple modes using simplex method
(dashed green lines).  The lighter lines denote the $1\sigma$ error
bars. \textit{Right panel}: zoom on $\l=3$ mode where an inequality breaks down
by $4.99\sigma$ (as indicated by the arrow).
\label{fig:summary}}
\end{figure}

The reason why the inequality failed for the $\l=3$ mode is straightforward to
understand.  The middle panel of Fig.~\ref{fig:simplex_model_S} shows the
resultant combined kernel, $\mathcal{K}$.  This kernel has a maximum amplitude
where the rotation profile increases sharply and a low amplitude near the sharp
decrease.  Hence, the integral of the product $-\dfrac{\Omega}{r} \mathcal{K}$
is negative which leads to a break-down of the inequalities for sufficiently
small error bars.  The lower panel of Fig.~\ref{fig:simplex_model_S} shows the
results obtained from a SOLA inversion of $\dfrac{\Omega}{x}$.  Given that only
the $\l=1$ to $3$ modes were used, the averaging kernels are poorly localised
and the resultant inverted profile a poor match to $\dfrac{\Omega}{x}$, which
inspires little confidence in the inversion.  Nonetheless, the
inversion does detect a positive gradient with a $2.87\sigma$ margin.  Hence,
the above inequalities, in conjunction with the simplex method, have yielded
firmer results than the inversion.

In contrast to the tests based on finding a positive gradient, those based on
Rayleigh's criterion failed to detect a decrease in the angular momentum. 
Possible explanations for this failure is that the region with this decrease is
either too narrow, or poorly located, given the set of modes we considered.

\begin{figure}[htbp]
\includegraphics[width=\columnwidth]{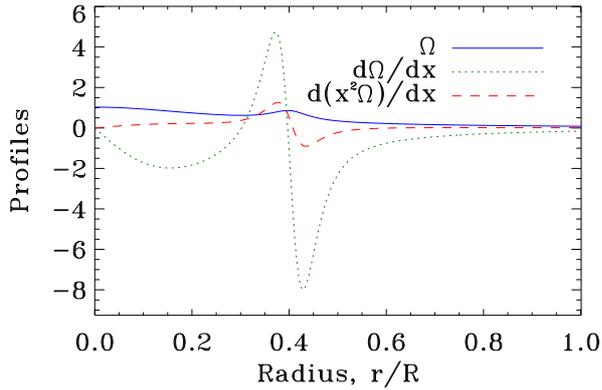}
\caption{(Colour online) The rotation profile, as based on
Eq.~(\ref{eq:rota_profile_forward}), and its
derivatives.\label{fig:rota_profiles_forward}}
\end{figure}

We also considered a second rotation profile with model S:
\begin{equation}
\Omega(x) = \frac{A}{1 + \left(\frac{x}{C}\right)^2} + \frac{B}{1 + \left(\frac{x^2-D^2}{E^2}\right)^2}
\label{eq:rota_profile_forward}
\end{equation}
where $x=r/R$, $A = 1\,\mu$Hz, $B = 0.5\,\mu$Hz, $C = 0.3$, $D=0.4$, and
$E=0.2$.  This rotation profile has been designed to have a positive gradient in
a localised region, fail Rayleigh's criterion in another region, and have a
zero derivative in the centre, as expected from regularity conditions. 
Figure~\ref{fig:rota_profiles_forward} illustrates this rotation profile,
as well as $\mathrm{d}\Omega/\mathrm{d}x$ and $\mathrm{d}(x^2
\Omega)/\mathrm{d}x$.

A set of $282$ modes with $\l$ ranging from $1$ to $20$ were calculated using
the ADIPLS pulsation code \citep{Christensen-Dalsgaard2008b}.  Such an extensive
set is naturally out of reach for stars other than the sun, given cancellation
effects, but is what is needed to carry out a sufficiently detailed inversion of
the rotation profile.   Rotational splittings were calculated using the above
profile, and their $1\sigma$ bars were set to $0.0033\,\mu$Hz,
which is smaller than what is currently achieved with the
\textit{Kepler} mission.

\begin{figure}[htbp]
\includegraphics[width=\columnwidth]{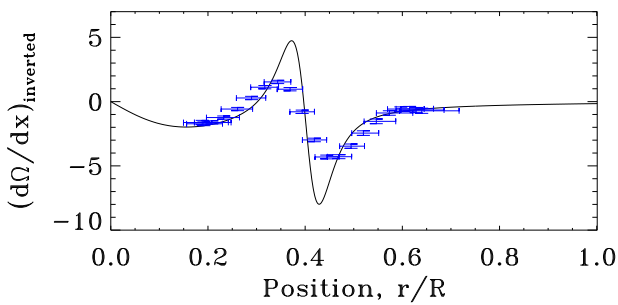}
\caption{(Colour online) Inverted rotation profile gradient.  The continuous
line shows the actual rotation profile gradient, whereas the blue crosses
indicate inversion results and associated $1\sigma$ error
bars.\label{fig:inversion}}
\end{figure}

Figure~\ref{fig:inversion} shows the result of a SOLA inversion
\citep{Pijpers1992, Pijpers1994} of the gradient of the rotation
profile\footnote{In order to cancel the surface term, we included the
supplementary constraint $K_{\mathrm{avg}}(R) = -\sum_i c_i I_i(R) = 0$ using a
Lagrange multiplier.  This can, however, lead to inversion coefficients which
are not as optimal for the above inequalities, since these only require
$\mathcal{K}(R) \geq 0$.}.  The $1\sigma$ error bars are displayed in
this plot.  The positive gradient shows up clearly.

In contrast to the inversions, a systematic application of the
inequalities~(\ref{eq:splittings_inequalities_final})
and~(\ref{eq:splittings_inequalities_no_centre_bound}) to all pairs of modes
did not reveal a positive gradient.  The same also applied to
tests based on multiple modes, as described above (we limited ourselves to modes
with $\l=1$ to $10$, given the numerical cost of the simplex algorithm). We also
applied a heuristic approach which consisted in recuperating the inversion
coefficients and adjusting them so as to have a positive combined kernel
everywhere.  This strategy was to no avail even if the error bars were
set to nearly 0. A possible explanation for these failures is the fact
that the rotation profile decreases only over a very narrow region.

\begin{figure}[htbp]
\includegraphics[width=\columnwidth]{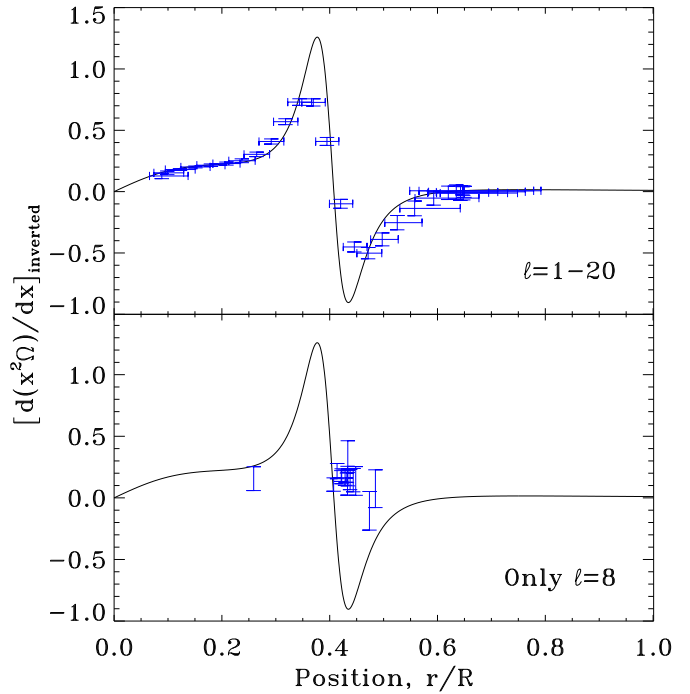}
\caption{(Colour online) Inverted profile for the gradient of the angular
momentum, using $\l=1$ to $20$ modes (upper panel) and  only $\l=8$ modes (lower
panel). The continuous line shows the actual gradient, whereas the blue crosses
indicate inversion results and associated $1\sigma$ error bars. For the sake of
clarity, the horizontal error bars (obtained from the averaging kernels) have
been omitted in the lower panel.\label{fig:momentum_inversion}}
\end{figure}

One may also see whether it is possible to show that the underlying profile does
not satisfy Rayleigh's criterion.  Figure~\ref{fig:momentum_inversion} shows
inversion\footnote{Once more, we removed the surface term by setting
$K_{\mathrm{avg}}(R) = -\sum_i c_i J_i(R) = 0$ through a Lagrange multiplier.}
results for the gradient of the angular momentum, using the integrated kernels
defined in Eq.~(\ref{eq:integrated_kernels_r2}).  As can be seen for the full
set of modes (upper panel), the inversion is able to detect the region which
does not satisfy Rayleigh's criterion.

\begin{figure}[htbp]
\includegraphics[width=\columnwidth]{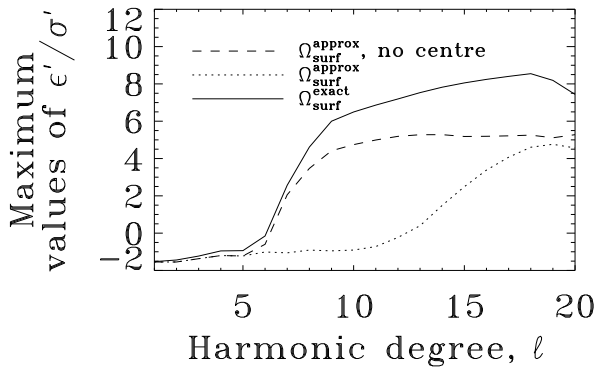}
\caption{Maximal values of $\varepsilon'/\sigma'$ as a function of $\l$
for inequalities based on Rayleigh's criterion.  Three different test cases are
investigated (see text for details). \label{fig:rayleigh}}
\end{figure}

We then systematically applied the
inequalities~(\ref{eq:splittings_inequalities_r2}) for each pair of modes.
Figure~\ref{fig:rayleigh} shows the maximal values of
$\varepsilon'/\sigma'$ as a function of the harmonic degree.  Three different
cases are investigated.  The solid line shows what happens if we use the exact
surface rotation rate, $\Omega(R) = 0.0836\,\mu$Hz.  Deviations bigger than
$4\sigma$ start to appear at $\l=8$.  The false alarm probability, based
on a Monte Carlo simulation is below $10^{-5}$ (\ie\ none of the $10^5$
realisations produced a large enough value of $\varepsilon'/\sigma'$).
The dotted line corresponds to the case 
where the overestimate $\Omega(R) = 0.5\,\mu$Hz is used.  This time, the
deviations exceeded $4\sigma$ for $\l\geq 17$.  The false alarm probability
is $1.4\times 10^{-3}$.  Finally, the dashed line shows
what happens when the centre is excluded, using $r_0 = 0.01$ and
$\tilde{\mathcal{B}}/R = 1.0\,\mu$Hz (which is nearly 50 times too large), and
still keeping $\Omega(R) = 0.5\,\mu$Hz.  Deviations greater than $4\sigma$
appear at $\l=9$.  The false alarm probability turned out to be
$1.3 \times 10^{-3}$.
We also carried out a Monte Carlo test with $10^{6}$ realisations in the
first case (\ie\ with the exact surface rotation rate, over the whole domain)
using only $\l=8$ modes. This yielded a false alarm probability of $2.1\times
10^{-4}$ (we note that the maximum value of $\varepsilon'/\sigma'$ is $4.61$ in this
case).   In contrast, an inversion based on just these modes yields rather poor
and inconclusive results due to poorly localised averaging kernels as
illustrated in the lower panel of Fig.~\ref{fig:momentum_inversion} (we do note
that one of the points in the inversion does suggest a decrease in the angular
momentum, but even the $1\sigma$ error bar allows for a positive gradient).

\begin{figure}[htbp]
\includegraphics[width=\columnwidth]{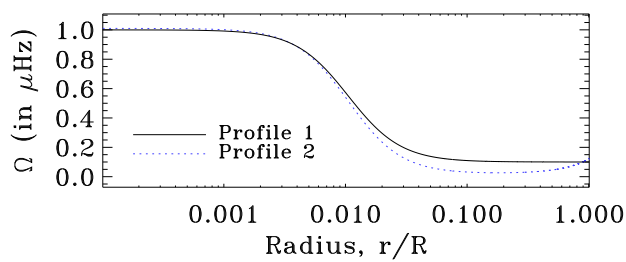}
\caption{(Colour online) Rotation profiles which give the similar splittings (to
within $0.01\,\mu$Hz) in Model 1 (from Sect.~\ref{sect:red_giants}).  Profile 1
is also used to calculate the ``observed'' rotational splittings (using Model 2)
in Sect.~\ref{sect:red_giants}.\label{fig:rota_profile}}
\end{figure}

One can also investigate the use of these inequalities in red giant stars. 
However, these seem to be less effective for such stars, since the rotation
kernels are mainly sensitive to the core or surface regions and hardly in
between.  Accordingly, it is not too difficult to construct pairs of rotation
profiles, one of which is decreasing the other not, and which produce
essentially the same rotational splittings (\ie\ to within some error bar). 
Figure~\ref{fig:rota_profile} shows such a pair -- the corresponding rotational
splittings, when calculated in Model~1 (or Model~2 -- these models are
introduced in the following section), are within $0.01\,\mu$Hz.  Indeed, as
explained in \citet{Goupil2013}, the rotational splittings of red-giants are
primarily sensitive to the average core and surface rotation rates.  Obviously,
the above inequalities will not detect a positive gradient in such a
situation, since there exists at least one decreasing rotation profile which
reproduces the splittings within the error bars. Including higher $\ell$ modes
doesn't seem like a viable solution, as these are harder to detect and will
experience a stronger trapping, either in the p- or g-mode cavities.  In order
to cause one of the inequalities to break down, one needs to have a
rotation profile which is increasing throughout most of the star.
However, this will probably cause the splittings of the p-like modes to
be comparable or larger than those of the g-like modes -- already a
very strong indication of an increasing rotation rate prior to any tests or
inversions.  In the following section, we nonetheless show how the above
inequalities can be useful when studying red giants.

\subsection{The effects of a mismatch between an observed star and the reference
model}
\label{sect:red_giants}

\begin{figure}[htbp]
\includegraphics[width=\columnwidth]{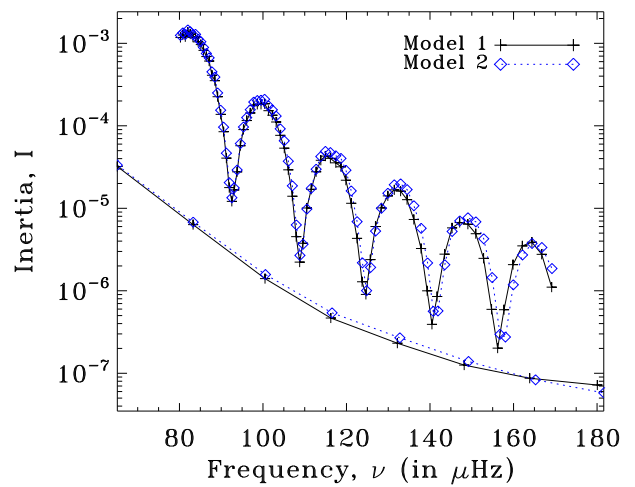}
\caption{(colour online) Inertias of $\l=0$ and $1$ modes in models~1 and~2
as a function of frequency.\label{fig:spectra}}
\end{figure}

At this point a natural question arises: if the above inequalities break
down is this necessarily caused by a positive gradient in the rotation profile,
or can there be other causes?  In what follows, we will show a situation where a
slight mismatch between an ``observed'' star and a reference model leads to one
of the inequalities breaking down, in spite of having a rotation profile where
the gradient is negative everywhere.  To do so, we start off with two very
close 1 M$_{\odot}$ models of red-giant stars. The second model is the
same as the first, except for an ad-hoc $50\%$ decrease of the density near the
surface. This is achieved by multiplying the density profile by the following
function:
\begin{equation}
f(r)=1+\frac{A\left\{1+\tanh\left[\frac{\lambda(r-r_d)}{R}\right]\right\}}{2},
\end{equation}
where $r_d = 0.995$, $\lambda=2\times 10^{3}$, and $A=-0.5$.
Table~\ref{table:models} gives the characteristics of these two models. We note
in passing that the mass difference between these two models corresponds to
$73\%$ of Mercury's mass. Figure~\ref{fig:spectra} shows the inertia of their
$\l=0$ and $1$ modes as a function of frequency, where $I_{n,\l}=\int_0^M
\|\vect{\xi}\|^2 \mathrm{d}m/[M\xi^2(R)]$. 

\begin{table}[htbp]
\begin{center}
\caption{Characteristics of the two models. The second column gives the
characteristics of Model~1, whereas the third column gives the relative
difference between Models~2 and~1.\label{table:models}}
\begin{tabular}{lcc}
\hline
\hline
\textbf{Quantity} &
\textbf{Model 1} &
\textbf{Relative difference} \\
\hline
Mass (in M$_{\odot}$) & $1.000$ & $-1.22 \times 10^{-7}$ \\
Radius (in R$_{\odot}$) & $4.107$ & $0$ \\
$\sqrt{GM/R^3}$ (in $\mu$Hz) & $12.00$ & $-6.1 \times 10^{-8}$ \\
\hline
\end{tabular}
\end{center}
\end{table}

\begin{figure}[htbp]
\includegraphics[width=\columnwidth]{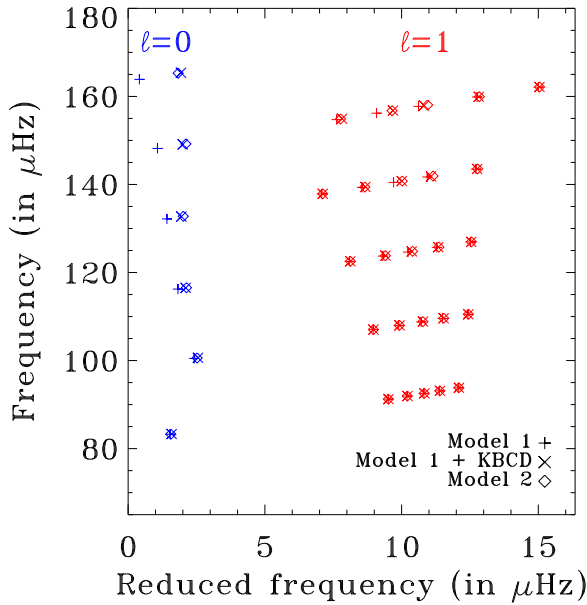}
\caption{Echelle diagram comparing the frequencies of Model 1 (with and without
the KBCD surface corrections) and Model 2.
\label{fig:echelle}}
\end{figure}

In what follows, the frequencies of Model 2 will be treated as the observations,
and Model 1 will be used as a reference model.  Furthermore, the ``observed''
frequency spectrum will include all of the $\l=0$ modes over a given frequency
range, and only a subset of the $\l=1$ modes.  Indeed, only the most p-like
dipole modes are expected to be visible in observed stars.  We therefore
selected the dipole modes at local minima of the inertia curve as well as the
adjacent 2 modes on either side.  Figure~\ref{fig:echelle} shows an echelle
diagram with the frequencies from both models.  We assume $1\sigma$
error bars of $0.01$ $\mu$Hz on the frequencies, which is typical for the
\textit{Kepler} mission. Table~\ref{tab:chi2} gives the value of
$\chi^2_{\mathrm{red}} = \frac{1}{M} \sum_{i=1}^{M}
\left(\frac{\nu^{\mathrm{ref}}_i - \nu^{\mathrm{obs}}_i}{\sigma_i}\right)^2$
where $M$ is the number of frequencies.  This quantity characterises the
departure of the frequency spectrum of Model 1 from that of Model 2.

\begin{table}[htbp]
\begin{center}
\caption{Seismic $\chi^2_{\mathrm{red}}$ values for Model 1, with and without
the KBCD surface corrections.\label{tab:chi2}}
\begin{tabular}{cc}
\hline
\hline
\textbf{Frequency set} &
$\chi^2_\mathrm{red}$ \\
\hline
Model 1 & 1364.5\\
Model 1 + KBCD & 19.4 \\
\hline
\end{tabular}
\end{center}
\end{table}

As can be seen from Table~\ref{tab:chi2}, the pulsation spectrum of Model 1 is a
poor fit to the observations.  Our observational error bars would therefore
exclude Model 1 immediately.   However, surface corrections based on
\citet{Kjeldsen2008} (KBCD, hereafter) are frequently included in seismic
studies of observed stars.  We therefore applied such corrections to the
frequencies of Model 1, using the following formula:
\begin{equation}
\nu_{n,\l}^{\mathrm{corrected}} = \nu_{n,\l} +
\frac{a_{\mathrm{KBCD}}\left(\frac{\nu_{n,\l}}{\nu_0}\right)^{b_{\mathrm{KBCD}}}}{Q_{n,\l}}
\end{equation}
where $\nu_{n,\l}$ are the original frequencies,
$\nu_{n,\l}^{\mathrm{corrected}}$ the corrected frequencies, $\nu_0$ a reference
frequency, $a_{\mathrm{KBCD}}$ a multiplicative constant, $b_{\mathrm{KBCD}}$ the
exponent used in KBCD's power law, and $Q_{n,\l}$ the mode's inertia, divided by
the inertia of an $\l=0$ mode interpolated to the same frequency
\citep[\eg][page 470]{Aerts2010}.  We choose the values $\nu_0 = 125\,\mu$Hz,
which is in the middle of the frequency range, $a_{\mathrm{KBCD}} = 0.40315$ in
order to reproduce as accurately as possible the frequency deviations in
Model~2, and $b_{\mathrm{KBCD}} = 4.9$, the value obtained in KBCD after
calibrating Model S to the sun.  The division by $Q_{n,\l}$ is a later addition
to the formula, which proved to be necessary in the case of red giants, due to
the large variations of the inertia of non-radial modes
\cite[\eg][]{Deheuvels2012}.  The second line of Table~\ref{tab:chi2} shows the
impact of including this surface correction.  Although the
$\chi^2_{\mathrm{red}}$ is still above $1$ and is worse than the best fitting
models in \citet{Deheuvels2012}, it is better than any of the best fits found in
\citet{Deheuvels2014}.  Hence, such a model would probably be retained as a
viable reference model with which to study the ``observed'' star (\ie\
Model~2).

\begin{figure}[htbp]
\includegraphics[width=\columnwidth]{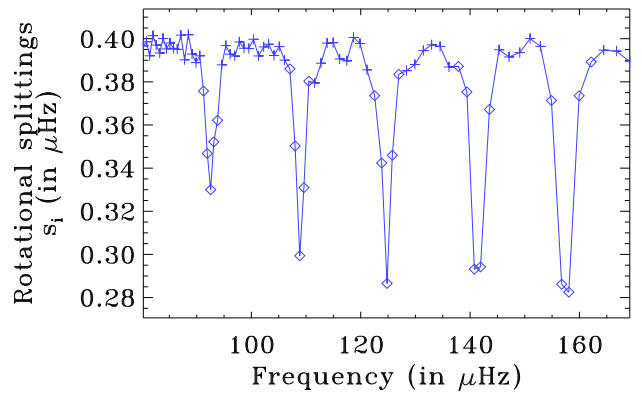}
\caption{Rotational splittings in model 2.  The diamonds correspond
to the splittings which were used in this study, whereas the crosses
represent those which were discarded. \label{fig:splittings}}
\end{figure}

We calculate rotational splittings in Model 2, using the following rotation
profile:
\begin{equation}
\Omega(r) = A + \frac{B}{1 + C(r/R)^2}
\label{eq:rotation_profile}
\end{equation}
where $A = 0.1$ $\mu$Hz, $B = 0.9$ $\mu$Hz and $C = 9 \times 10^3$.  We note
that this rotation profile decreases with $r$, but still satisfies Rayleigh's
stability criterion.  This rotation profile is illustrated in
Fig.~\ref{fig:rota_profile} (see Profile 1).  The resultant rotational
splittings for model 2 are shown in Fig.~\ref{fig:splittings}, and these will
serve as our observations.  We assume $1\sigma$ error bars of
$0.007$ $\mu$Hz on these splittings, which is approximately
$\sqrt{2}$ times smaller than the error bars on the frequencies, which is
what you would expect from generalised $\l=1$ splittings.

\begin{figure}[htbp]
\includegraphics[width=\columnwidth]{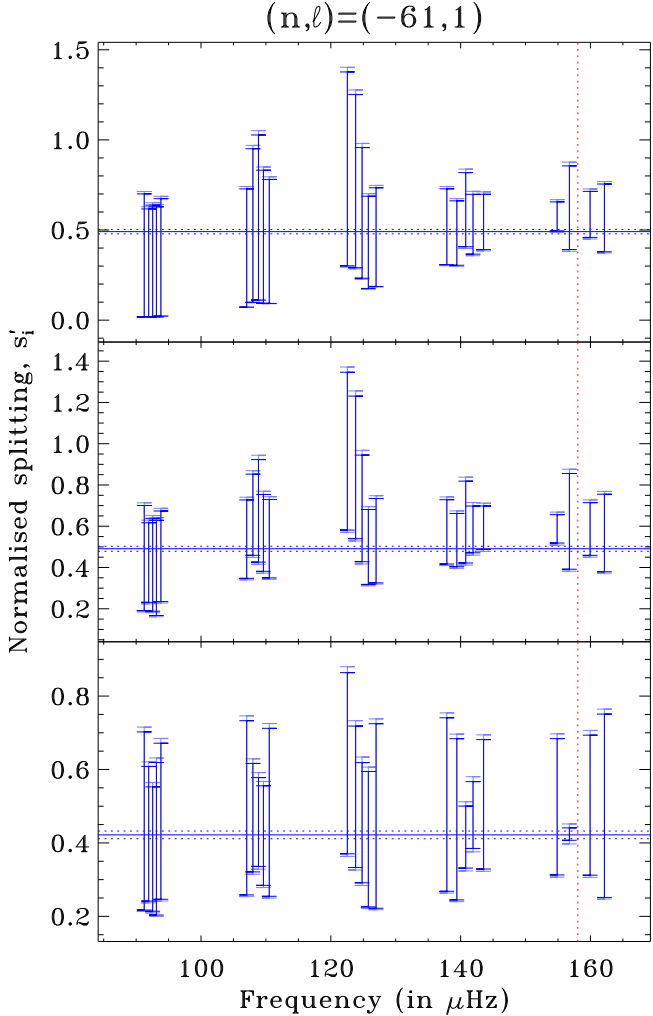}
\caption{(colour online) Upper and lower bounds on a given splitting, deduced
from the $N-1$ other rotational splittings, assuming that the rotation profile
decreases.  In the upper and middle panels, Model~1 has been used as the
reference model, whereas in the lower panel, Model~2 is the reference model.  In
the lower two panels, a more restrictive set of inequalities is obtained by
excluding the centre, as described in Sect.~\ref{sect:inequalities_no_centre}. 
The frequency of the original mode is shown by the red vertical dotted line,
whereas the blue horizontal dotted and solid lines indicate its rotational
splitting plus or minus the error bar.  The dark blue solid vertical
segments indicate the upper and lower bounds deduced from the other modes,
and the $1\sigma$ error bars are indicated by the light blue
extensions.\label{fig:splittings_mismatch}}
\end{figure}

Then, we systematically apply Eq.~(\ref{eq:splittings_inequalities_final}) to
every pair of dipole modes. For each mode, we end up with $N-1$ upper and lower
bounds, where $N$ is the number of rotational splittings.
Figure~\ref{fig:splittings_mismatch} (upper panel) shows the bounds which are
obtained for one of the p-like modes in Model 1. According to this approach,
this splitting is compatible with the lower and upper bounds from the other
modes to within the $1.22\sigma$. However, if the centre is excluded,
as explained in Sect.~\ref{sect:inequalities_no_centre}, then it is possible to
find an inequality which breaks down by $5.87\sigma$.  The middle panel
of Figure~\ref{fig:splittings_mismatch} shows the bounds obtained for the same
mode, but using Eq.~\ref{eq:splittings_inequalities_no_centre_bound} with $r_0 =
2 \times 10^{-4} R$ and $R \mathcal{B} = 5 \times 10^{5}\,\mu$Hz.  We note that
the actual upper bound over the interval $[0,r_0]$ is $R \mathcal{B} =
3.270\,\mu$Hz, \ie\ more than 5 orders of magnitude smaller.  A Monte-Carlo
simulation based on the coefficients deduced from inequalities on mode pairs 
found three cases where $\varepsilon'/\sigma' > 5.87$ in
$10^6$ realisations, thereby indicating a false alarm probability of
$3 \times 10^{-6}$.  The lower panel shows what happens when applying
the same tests using Model~2 as a reference model.  This time we set $R
\mathcal{B} = 3.270\,\mu$Hz to be as restrictive as possible.  Even then, none
of the inequalities break down, which is consistent with the fact that we are
using the true model and the rotation profile is decreasing.

\begin{figure}[htbp]
\includegraphics[width=\columnwidth]{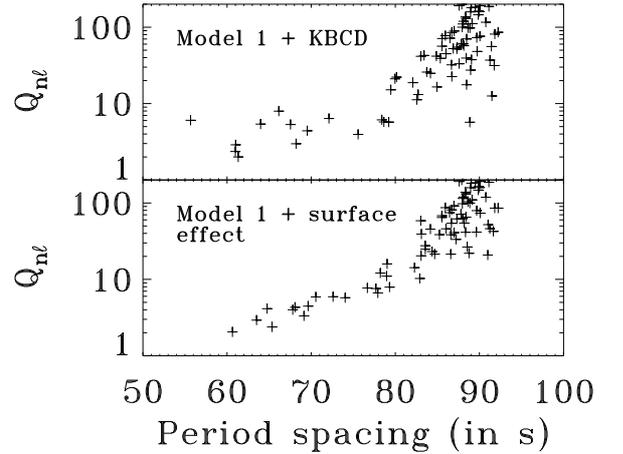}
\caption{Normalised mode inertias as a function of averaged periods spacings
(defined as $\pi_n = (1/\nu_{n-1} - 1/\nu_{n+1})/2$, where $n$ is the radial
order).  The lower panel corresponds to Model~1 with a $90\,\%$ decrease
in density near the surface, whereas the upper panel uses Model~1 with KBCD
surface corrections, adjusted so as to match (as best as possible) the
frequencies of the modified model.  The increased scatter at low inertias
in the upper panel illustrates the mismatch between period spacings and mode
inertias.\label{fig:limits_of_KBCD}}
\end{figure}

It is relatively straightforward to understand why Model~1 leads to a break down
of one of the inequalities.  Indeed, as can be seen in Fig.~\ref{fig:spectra},
the mode inertias of Models~1 and~2 do not match very well, especially for the
high frequency p-like dipole modes.  This modification of mode inertias stems
from the fact that surface effects act differently on p and g-modes and will
therefore shift the former with respect to the latter.  This, in turn, modifies
the avoided crossings which occur between them, and hence, the inertias of the
resultant modes. Such an effect is likely to be strongest in red giant stars
where the spacing between consecutive g-modes can be comparable to the frequency
shifts caused by surface effects.  In contrast, applying a surface correction
recipe like the one described in KBCD merely corrects the frequencies and not
the inertias.  This leads to inconsistencies between the mode inertias and
period spacings, as illustrated in Fig.~\ref{fig:limits_of_KBCD}, and could mask
mismatches between observations and what seems to be the best-fitting model(s).

\begin{figure}[htbp]
\includegraphics[width=\columnwidth]{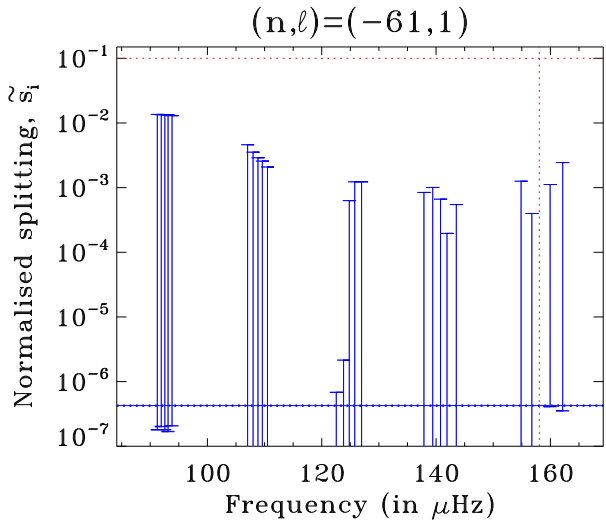}
\caption{(colour online) Same as Fig.~\ref{fig:splittings_mismatch}
but using the inequalities based on Rayleigh's criterion.  The centre
has also been excluded, as described in the text, thereby leading to
more stringent conditions. The horizontal red dotted line near the
top of the figure is an upper limit deduced from the surface rotation
rate.  The error bars which extend off the bottom of the plot go to
negative values.\label{fig:splittings_mismatch_r2}}
\end{figure}

One can also apply the tests based on Rayleigh's criterion.  We therefore
systematically applied Eq.~(\ref{eq:splittings_inequalities_r2}) using the exact
surface rotation rate, to every pair of modes.  However, this set of
inequalities turns out to be far less stringent than the first set of
inequalities.  Accordingly, none of them break down, even when the centre is
excluded, as described in Sect.~\ref{sect:inequalities_r2_no_centre}, using the
most restrictive value of $\tilde{\mathcal{B}}$ possible. 
Figure~\ref{fig:splittings_mismatch_r2} shows lower and upper bounds for the
rotational splitting of the dipole mode $(n,\l)=(-61,1)$ with $r_0/R = 2\times
10^{-4}$ and $\tilde{\mathcal{B}}/R = 4.037\times 10^{-4}\,\mu$Hz.

One could then turn the inequalities based on Rayleigh's criterion the
other way around and search for a lower bound on the surface rotation rate.
Equation~(\ref{eq:inequalities_r2_simple}) already provides a first condition. 
Applying this condition using Model 1 as the reference model leads to
$\Omega(R)>6.54 \times 10^{-7}\,\mu$Hz whereas Model 2 (the true model) leads
to $\Omega(R) >6.04 \times 10^{-7}\,\mu$Hz. The true surface rotation rate is
more than 5 orders of magnitude larger, with $\Omega(R) = 0.10010\,\mu$Hz. 
However one can also rearrange
Eq.~(\ref{eq:splittings_inequalities_r2_no_centre}) to obtain a more restrictive
condition:
\begin{eqnarray}
\frac{\btbis \tilde{s}_j - \tilde{s}_i - \tilde{\mathcal{B}} \!\! \int_0^{r_0}
\!\! \max \left[0, J_i(r) - \btbis J_j(r)\right] \mathrm{d} r}
{R^2(\btbis - 1)} &\leq& \Omega(R) \\
\frac{\tilde{s}_i - \atbis \tilde{s}_j - \tilde{\mathcal{B}} \!\! \int_0^{r_0}
\!\! \max \left[0, \atbis J_j(r) - J_i(r)\right] \mathrm{d} r}
{R^2(1 - \atbis)} &\leq& \Omega(R)
\end{eqnarray}
Applying these conditions using $r_0 = 2 \times 10^{-4}$ and
$\tilde{\mathcal{B}}/R = 4.037 \times 10^{-4}\,\mu$Hz yields $\Omega(R) \geq
4.89\times 10^{-4}\,\mu$Hz for Model~1 and $\Omega(R) \geq 1.52 \times
10^{-5}\,\mu$Hz for Model~2, which is still 3 to 4 orders of magnitude below the
true value, and furthermore, very model dependant.  Hence, even this way around,
the inequalities based on Rayleigh's criterion only provide very weak constraints
and one would probably be better off either trying to measure the rotation rate
through photometric modulation induced by spots (if present) or through
spectroscopy.  Alternatively, one could apply the relation derived by
\citet{Goupil2013} which relates the ratio between the average surface and core
rotation rates of red giants to the slope of the rotational splittings as a
function of $\zeta$, the ratio of the kinetic energy of the mode in the g-mode
cavity to the total kinetic energy.  Such a relation has already been tested on
$6$ subgiants observed by Kepler and shown to agree, in most cases, with results
based on inversions \citep{Deheuvels2014}.

\section{Discussion}

In this article, we have derived various inequalities which can be used to check
whether a rotation profile has a negative gradient or satisfies Rayleigh's
stability criterion, for a given reference model.  We explored how these
inequalities may be used in practical situations by investigating various test
cases.  In the first test case, a rotation profile which satisfies neither the
assumption $\dfrac{\Omega}{r} \leq 0$ everywhere nor Rayleigh's criterion was
imposed on Model S \citep{Christensen-Dalsgaard1996}. A set of $44$
splittings for modes with $\l=1$ to $3$ were used in the various tests. The
above inequalities were able to detect the presence of a positive gradient, but
were unable to show a violation of Rayleigh's criterion.  Although an inversion
also suggested the presence of a positive gradient, the resultant conclusions
remained less firm than those derived by the inequalities. A second test case,
also based upon Model S, used a different profile which also contained a region
positive gradient on the rotation profile and one where the angular momentum is
decreasing.  This second case used a set of $282$ modes (with $\l=1$ to $20$), 
thereby enabling a clear detection of both of these regions using inversions. 
In contrast, the above inequalities failed to detect the positive gradient in
the rotation profile, probably as a result of the narrowness of this region. 
However, these were able to detect a violation of Rayleigh's criterion, even in
situations where inversions yielded poor results, provided modes with $\l \geq
8$ were included.   We expect that this lower limit on the $\l$ value is
dependant on both the way in which the rotation profile violates Rayleigh's
criterion and on the size of the error bars on the rotational splittings.

We also looked at other ways in which the inequalities may break down. 
For this purpose, we considered two models of red-giants which were identical
except for an ad-hoc near-surface modification of the density profile in the
second model.  The first model was then used as a reference model in trying to
interpret the rotational splittings of the second model, which were based on a
rotation profile which has a negative gradient everywhere and which obeys
Rayleigh's criterion.  It turned out that one of the inequalities based on the
rotation profile's gradient was violated, whereas all of the inequalities based
on Rayleigh's criterion were satisfied.  The reason for this is that the mode
inertias differ in the two models, especially at high frequencies.  This may
come as a surprise since only the outer $0.5\,\%$ or so is different between the
two models.  However, as argued above, near-surface effects shifts the
frequencies of p-like modes relative to those of the g-like modes, thereby
modifying the avoided crossings which occur between the two, and hence the
resultant inertias.  Frequency correction recipes for near-surface effects, such
as the one presented in \citet{Kjeldsen2008}, are able to mask the frequency
differences between the two models, but they do not correct the mode inertias,
thereby leading to erroneous conclusions on the rotation profile as shown in
this test case.

This leads to the natural question as to what to do if one or several
inequalities are violated by a set of observed rotational splittings. 
In answer to this question, one needs to consider the following
options.  First, a different, more suitable, reference model needs to be found. 
We expect that in the case of red giants, the above inequalities could be quite
constraining, due to very different mode trappings, and may therefore provide an
indirect way of testing the structure of the reference model. Furthermore, such
stars are the most likely candidates to having a rotation profile which
decreases outward due to the spin-up of the core as it contracts and the slowing
down of the envelope as it expands.  Of course, a direct comparison of
the non-radial mode frequencies will very likely provide tighter constraints,
but so far, no one has yet fitted a red-giant model to within $1\sigma$ of a set
observed frequencies.  Hence, if one adopts less stringent constraints on the
selection of the model, and/or if frequency surface-corrections are included,
the above criteria will be useful in determining beforehand whether or not the
chosen reference model is compatible with a decreasing rotation profile.  The
second option, is to consider that the rotation profile doesn't have a negative
gradient throughout the star and/or violates Rayleigh's stability criterion
(depending on which inequalities are not satisfied).  As was already stated in
the introduction, there is convincing evidence of stars where the gradient of
the rotation profile is positive, at least in localised regions
\citep{Schou1998, Kurtz2014}.  This raises the open questions as to how often
such a situation occurs, and whether even a red giant couldn't have a localised
region with a positive rotation gradient, for example, in the outer convective
envelope.  However, one must not forget that the above criteria are not
necessarily the most effective at detecting a rotation profile with a positive
gradient. Accordingly, a mismatch between the reference model and the observed
star may be a more likely explanation, which brings us back to the first
option.  A last option would be to consider that the rotation profile is too
rapid to be described by first order effects, thereby violating the assumptions
in this work.  In such a situation, it would no longer be possible to apply
linear inversion theory, except perhaps on generalised rotational splittings
(defined as the frequency difference between prograde modes and their retrograde
counterparts) of acoustic modes in stars with nearly uniform rotation profiles
\citep{Reese2009a}.  Recent works \citep{Reese2006, Ballot2010, Ouazzani2013}
have, however, shown up to what point a first order description of the effects
of rotation are valid, and can therefore be used to decide whether or not the
criteria described in this paper are applicable.

\begin{acknowledgements}
DRR thanks the referee for very constructive comments which have substantially
improved the manuscript. DRR thanks A. Miglio for helping him come up with some
of the ideas in this article and for providing red giant models. DRR is
currently funded by the European Community's Seventh Framework Programme
(FP7/2007-2013) under grant agreement no. 312844 (SPACEINN), which is gratefully
acknowledged. This article made use of InversionKit, an inversion software
developed in the context of the HELAS and SPACEINN networks, funded by the
European Commission's Sixth and Seventh Framework Programmes.
\end{acknowledgements}

\bibliographystyle{aa}
\bibliography{biblio}

\appendix

\section{Attempting to construct a decreasing rotation profile when
Eq.~(\ref{eq:splittings_inequalities_no_centre_bound}) is satisfied}
\label{sect:solution_no_centre}

We start by introducing the following radial coordinates:
\begin{equation}
r_{\abis} = \argmin_{r \in [r_0,R]} \left(\frac{I_i(r)}{I_j(r)}\right), \qquad
r_{\bbis} = \argmax_{r \in [r_0,R]} \left(\frac{I_i(r)}{I_j(r)}\right).
\end{equation}
and define a profile as follows (where we have assumed $r_{\abis} <
r_{\bbis}$):
\begin{equation}
\Omega(r) = \left\{
\begin{array}{ll}
\lambda + \mu \qquad & \mbox{ if } r_0 \leq r \leq r_{\abis} \\
\mu  & \mbox{ if } r_{\abis} < r \leq r_{\bbis} \\
0 & \mbox{ if } r_{\bbis} < r \leq R
\end{array}
\right..
\end{equation}
A similar solution can be obtained if $r_{\bbis} < r_{\abis}$.  In this
definition, we have deliberately postponed defining the rotation profile over
the interval $[0, r_0[$, except that we will impose its continuity at $r_0$
(except if $r_0 = r_{\abis}$).  Inserting this expression into
Eq.~(\ref{eq:inverse_problem_reformulated_no_centre}) then yields the following
system:
\begin{eqnarray}
s'_i + \int_0^{r_0} \dfrac{\Omega}{r} I_i(r) \mathrm{d}r &=& \lambda I_i(r_{\abis}) + \mu I_i(r_{\bbis}), \\
s'_j + \int_0^{r_0} \dfrac{\Omega}{r} I_j(r) \mathrm{d}r &=& \lambda I_j(r_{\abis}) + \mu I_j(r_{\bbis}),
\end{eqnarray}
the solution of which is:
\begin{eqnarray}
\label{eq:lambda_bis}
\lambda &=& \frac{1}{I_j(r_{\abis})} \left\{ \frac{\bbis s'_j - s'_i + \int_0^{r_0} \dfrac{\Omega}{r}
            \left[\bbis I_j(r) - I_i(r)\right] \mathrm{d} r}{\bbis-\abis} \right\}, \\
\label{eq:mu_bis}
\mu     &=& \frac{1}{I_j(r_{\bbis})} \left\{ \frac{s'_i - \abis s'_j + \int_0^{r_0} \dfrac{\Omega}{r}
            \left[I_i(r) - \abis I_j(r)\right] \mathrm{d} r}{\bbis-\abis} \right\}.
\end{eqnarray}
From this we deduce that
Eq.~(\ref{eq:splittings_inequalities_no_centre}) is a sufficient condition to
ensure that $\lambda$ and $\mu$ are positive, and hence that $\Omega$  is
decreasing.  In contrast, Eq.~(\ref{eq:splittings_inequalities_no_centre_bound}),
does not ensure that $\lambda$ and $\mu$ are positive.  To see this, we re-express
the numerator of Eq.~(\ref{eq:lambda_bis}) as follows:
\begin{eqnarray*}
&   & \underbrace{\bbis s'_j - s'_i - \int_0^{r_0} \dfrac{\Omega}{r} \max\left[0,I_i(r) - \bbis  I_j(r)\right] \mathrm{d} r}_{I} \\
&   & \underbrace{-\int_0^{r_0} \dfrac{\Omega}{r} \min\left[0,I_i(r) - \bbis  I_j(r)\right] \mathrm{d} r}_{II}.
\end{eqnarray*}
In order to ensure that the term $I$ is positive according to
Eq.~(\ref{eq:splittings_inequalities_no_centre_bound}), one needs to set
$\dfrac{\Omega}{r}$ to $-\mathcal{B}$ over the interval where $I_i(r) - \bbis
I_j(r)$ is positive. Term $II$ is necessarily smaller or equal to zero.  In
order to cancel it out, $\dfrac{\Omega}{r}$ needs to be zero over the interval
where $I_i(r) - \bbis I_j(r)$ is negative.  Likewise, in order to ensure that
$\mu$ is positive, one would need to make sure $\dfrac{\Omega}{r} =
-\mathcal{B}$ when $\abis I_i(r) - I_j(r)$ is positive, and $\dfrac{\Omega}{r} =
0$ otherwise.  However, there is no guarantee that these two sets of conditions
are compatible.

\end{document}